\documentclass[twocolumn,showpacs,aps,prd,nobibnotes,nofootinbib,floatfix]{revtex4-1}

\usepackage{amsmath}
\usepackage{graphicx,subfigure}
\usepackage[usenames]{color}
\usepackage[normalem]{ulem}
\usepackage{textcomp}
\usepackage{gensymb}
\usepackage{xspace}
\usepackage{verbatim}

\usepackage{soul}

\usepackage[colorlinks=true]{hyperref}
\hypersetup{
  citecolor  = blue
}
\newcommand{\rh}{r_{\text{h}}}
\newcommand{\re}{r_{\text{ex}}}
\newcommand{\GL}{G_{\text{L}}}

\begin{document}

\title{Kinetics of a phase transition for a Kerr-AdS black hole on the free-energy landscape}

\author{Si-Jiang Yang$^a$$^b$}
\author{Run Zhou$^a$$^b$}
\author{Shao-Wen Wei$^a$$^b$}

\author{Yu-Xiao Liu$^a$$^b$$^c$}%
\email{liuyx@lzu.edu.cn, corresponding author}

\affiliation{$^{a}$Lanzhou Center for Theoretical Physics, Key Laboratory of Theoretical Physics of Gansu Province, School of Physical Science and Technology, Lanzhou University, Lanzhou 730000, China\\
$^{b}$Institute of Theoretical Physics $\&$ Research Center of Gravitation, Lanzhou University, Lanzhou 730000, China\\
$^{c}$Key Laboratory for Magnetism and Magnetic of the Ministry of Education, Lanzhou University, Lanzhou 730000, China}

\date{\today}

\begin{abstract}
By treating the order parameter as a stochastic thermal fluctuating variable for small-large black hole phase transition, we investigate the kinetic process of phase transition for the Kerr-AdS (anti-de Sitter) black holes on free energy landscape. We find that the extremal points of the off shell Gibbs free energy correspond to physical black holes. For small-large black hole phase transition, the off shell Gibbs free energy exhibits a double well behavior with the same depth. Contrary to previous research for the kinetics of phase transition for the Kerr-Newman-AdS family black holes on a free energy landscape, we find that there is a lower bound for the order parameter and the lower bound corresponds to extremal black holes. In particular, the off shell Gibbs free energy is zero instead of divergent as previous work suggested for vanishing black hole horizon radius, which corresponds to the Gibbs free energy of a thermal AdS space. The investigation for the evolution of the probability distribution for the phase transition indicates that the initial stable small (large) black hole tends to switch to stable large (small) black hole. Increasing the temperature along the coexistence curve, the switching process becomes faster and the probability distribution reaches the final stationary Boltzmann distribution at a shorter time. The distribution of the first passage time indicates the time scale of the small-large black hole phase transition, and the peak of the distribution becomes sharper and shifts to the left with the increase of temperature along the coexistence curve. This suggests that a considerable first passage process occurs at a shorter time for higher temperature. The investigation of the kinetics of phase transition might provide us new insight into the underlying microscopic interactions.
\end{abstract}

\maketitle

\section{Introduction}\label{Sec:Intro}

Black hole thermodynamics has remained an intriguing subject and attracted great interest for many decades. This is largely due to the fact that black hole thermodynamics has brought to light strong hints of a very deep and fundamental connection between gravitation, thermodynamics, and quantum theory~\cite{Wald01}. For example,  Einstein field equations can be derived from thermodynamics and can be viewed as an equation of state~\cite{Jaco95}. Research of black hole thermodynamics suggests that black holes cannot only be assigned standard thermodynamic variables, such as temperature and entropy~\cite{Beke72,Beke73,Hawk74,Hawk75}, but also might have microscopic structure~\cite{Stro98,WeLi15,WeLM19,WeLi20}. The connections between gravity and other branches of theoretical physics have brought new insight into the development of gravitation but also aroused puzzling questions~\cite{Padm10}.

It is known that there are various phase transitions for black holes in anti-de Sitter (AdS) space. Particularly interesting are the Hawking-Page phase transition, the small-large black hole phase transition and the triple point phase transition. The Hawking-Page phase transition between the large stable Schwarzschild-AdS black hole and radiation~\cite{HaPa83} can be interpreted as a confinement/deconfinement phase transition in the dual quark gluon plasma~\cite{Witt98}. Phase transition for the charged AdS black hole shows great resemblance to that of the Van der Waals fluid~\cite{CEJM99a,CEJM99}. In fact, the analogy turns out to be comparing the ``same physical quantities'' after the interpretation of the cosmological constant as thermodynamical pressure~\cite{KaRT09}, since they have the same critical exponents and the same universal number~\cite{Dola11,KuMa12}. Further investigation of other asymptotically AdS black holes in extended phase space shows that they have phase structure similar to that of ``every day thermodynamics'' and have been intensively studied in a broad variety of contexts~\cite{YaHD18,WeLi13,LLML17,CCLY13,ZoLW14,XuWY21,MiXu18,MiXu19,HuZL13,ArGG21}. The behavior of phase transition for high-dimensional multiple rotating Kerr-AdS black holes displays a small/intermediate/large black hole phase transition with one triple point and two critical points, which is reminiscent of the solid/liquid/gas phase transition~\cite{AlKM13,AKMS14}. There even exists a phenomenon of multiple reentrant phase transitions for asymptotically AdS black holes in Lovelock gravity and quasitopological gravity~\cite{FKMS14,HeBM15}. Phase transition for a broad class of asymptotically AdS hairy black holes in Lovelock gravity strongly resembles those of superfluidity in liquid helium~\cite{HeMT17}. Further investigation reveals that properties of phase transition for asymptotically AdS black holes can be reflected from the quasinormal modes and null geodesics~\cite{LiZW14,WeLW19,XWLW19}. For a comprehensive review of black hole thermodynamics and phase transition in extended phase space, see Refs.~\cite{AKMS14a,KuMT17}.

Compared with the equilibrium thermodynamics, studying the kinetic process of phase transition is a challenging issue. Inspired by the study of protein folding on the free energy landscape~\cite{FrSW91,BrWo89}, where a protein folding process can be regarded as the phase transition of states depicted by the stochastic Fokker-Planck equation~\cite{Risk84}, Li and Wang proposed a heuristic treatment of the kinetic process of black hole phase transition based on the free energy landscape recently~\cite{LiWa20,LiZW20}. The stochastic treatment of the Fokker-Planck equation towards describing the distribution of stochastic physical variables has been widely used in chemistry, biology, physics, and astrophysics~\cite{LLSW03,LeSW03,Wang15,Chan43,Amar18}. From the free energy landscape perspective, the black hole event horizon is the order parameter, and black hole phase transition is regarded as a stochastic thermal fluctuating of the order parameter~\cite{LiZW21}. The evolution of the probability distribution for the order parameter is governed by the Fokker-Planck equation.

Based on this viewpoint, the kinetic process of the Hawking-Page phase transition was investigated both in Einstein gravity and in massive gravity. It was found that the system has a probability changing from a black hole to a thermal AdS space, and vice versa~\cite{LiWa20,LiZW21a}. This was generalized to the investigation of the kinetic process of phase transition for charged AdS black hole, where the switching process between a metastable black hole and a stable black hole was considered~\cite{LiZW20}. Through numerical evaluation of the Fokker-Planck equation, the results suggest that the initial small black hole tends to switch to the large black hole, and the reverse process can also occur.
The investigation of the five-dimensional neutral Gauss-Bonnet-AdS black hole for kinetic process of phase transition on free energy landscape demonstrates that physical black holes correspond to the extremal points of the off shell Gibbs free energy, and the barrier height for the stable small-large black hole phase transition decreases with the increase of temperature~\cite{WeLW20}. Moreover, research of the dynamical properties of phase transition on free energy landscape for the six-dimensional charged Gauss-Bonnet-AdS black hole demonstrates that the triple point exhibits interesting oscillatory behavior~\cite{WWLM21,Cai:2021sag}, and the investigation of the  effect of quintessence dark energy on the kinetics of black hole phase transition indicates the larger the state parameter of
quintessence dark energy the faster the black hole system evolves~\cite{LMLX21}. Further investigation of thermodynamics on free energy landscape suggests that it might uncover the underlying microscopic interactions of black holes~\cite{LiWa20a,LiZW21}.

From the free energy landscape perspective, the off shell Gibbs free energy plays the role of an effective potential and acts as a driving force for black hole phase transition~\cite{LiZW21}. Hence, the behavior of the off shell Gibbs free energy is vital for describing the kinetics of black hole phase transition. While previous research for the kinetics of phase transition for charged AdS black hole on the free energy landscape argues that the off shell Gibbs free energy can be regarded as a function of the black hole horizon radius $\rh$, charge $Q$, thermodynamic pressure $P$ and the temperature of the ensemble $T_{\text{E}}$, and the black hole horizon radius can take any positive value. In particular, previous research argues that the off shell Gibbs free energy is divergent when the black hole horizon radius approaches zero~\cite{LiZW20}. However, this contradicts the physical intuition that vanishing black hole event horizon radius corresponds to a thermal AdS space. The enthalpy and entropy of a thermal AdS space are zero. Hence, the off shell Gibbs free energy should be zero. Furthermore, the thermal AdS space is also a solution of Einstein field equations, it should be a point on the off shell Gibbs free energy. On the other hand, thermodynamics along the coexistence curve for a black hole system is also an interesting subject.

Motivated by these, we investigate the behavior of the off shell Gibbs free energy for the Kerr-AdS black holes and explore the kinetics of the small-large black hole phase transition by increasing the temperature along the coexistence curve. We find that there is a lower bound for the order parameter and the lower bound corresponds to extremal black holes. Moreover, the off shell Gibbs free energy is zero instead of divergent as previous work indicates for vanishing black hole horizon radius, and it corresponds to the thermal AdS space. Our research is consistent with physical intuition that vanishing horizon radius corresponds to a thermal AdS space. The investigation of the kinetic process of small-large black hole phase transition suggests that the evolution process becomes quicker and reaches the final stationary Boltzmann distribution at a shorter time with the increase of the temperature along the coexistence curve.


The outline of the paper is as follows. In Sec.~\ref{Sec:thermo}, we briefly review the thermodynamics and coexistence curve for the Kerr-AdS black hole, and give a brief discussion on the order parameter for the small-large black hole phase transition. In Sec.~\ref{Sec:lands}, after a brief discussion of the off shell Gibbs free energy, we argue that there should be a lower bound for the order parameter for black hole phase transition, and the lower bound corresponds to extremal black holes. In Sec.~\ref{Sec: Dynmic}, we explore the kinetics of the small-large black hole phase transition for the Kerr-AdS black hole by solving the Fokker-Planck equation numerically, and investigate the effects of the temperature on the kinetic process of the small-large black hole phase transition by increasing the temperature of the ensemble along the coexistence curve. The last section is devoted to discussion and conclusion.

\section{Extended phase space thermodynamics and coexistence curve}\label{Sec:thermo}

The Kerr-AdS black hole is a four-dimensional rotating AdS black hole, which was first constructed by Carter~\cite{Cart68}. The thermodynamics and phase structure of the Kerr-AdS black hole in extended phase space exhibit the Van der Waals-like phase transition~\cite{AlKM13,AKMS14}, and the small-large black hole coexistence curve and critical point have been obtained in Refs.~\cite{WeCL16,GuKM12}. In this section, we briefly review the thermodynamics and coexistence curve for the Kerr-AdS black hole.

\subsection{Thermodynamics of Kerr-AdS black hole}\label{Sec:thermodynamics}

The Kerr-AdS black hole is a vacuum solution of Einstein field equations with a negative cosmological constant.
In Boyer-Lindquist coordinates $(t, r, \theta, \phi)$, the metric for the Kerr-AdS spacetime can be written as
\begin{multline}
  ds^2 = -\frac{\Delta}{\rho^2}\left(dt-\frac{a\sin^2\theta}{\Xi}d\phi\right)^2+ \frac{\rho^2}{\Delta}dr^2 \\
   + \frac{\rho^2}{\Sigma}d\theta^2+ \frac{\Sigma\sin^2\theta}{\rho^2}\left(adt-\frac{r^2+a^2}{\Xi}d\phi\right)^2,\label{KAdS}
\end{multline}
with the metric functions
\begin{align}
\notag \Delta & =(r^2+a^2)\left(1+\frac{r^2}{l^2}\right)-2mr, &\quad  \Xi & =1-\frac{a^2}{l^2} , \\
   \rho^2&=r^2+a^2\cos^2\theta, &\quad \Sigma &=1-\frac{a^2}{l^2}\cos^2\theta,
\end{align}
where $m$, $a$, and $l$ are the mass parameter, rotation parameter, and AdS radius, respectively.

The event horizon radius $r_{\text{h}}$ of the black hole is determined by the equation
\begin{equation}\label{horizonD}
  \Delta = (r^2+a^2)\left(1+\frac{r^2}{l^2}\right)-2mr=0.
\end{equation}
The Kerr-AdS metric describes a rotating spacetime with the following angular velocity
\begin{equation}\label{angul}
  \Omega_{\text{H}}=\frac{a}{l^2}\frac{r^2_{\text{h}}+l^2}{r^2_{\text{h}}+a^2}.
\end{equation}
In the extended phase space, the mass of the black hole is interpreted as enthalpy instead of internal energy~\cite{KaRT09}. The mass and angular momentum of the black hole were first obtained from the Hamiltonian approach by considering the generators of $SO(3,2)$~\cite{GiPP05,HeTe85}:
\begin{equation}
   M=\frac{m}{\Xi^2},  \qquad J=\frac{ma}{\Xi^2},
\end{equation}
and they are related to the mass parameter $m$ and angular parameter $a$, respectively.
The Hawking temperature of the black hole is
\begin{equation}\label{HawkTemp}
  T_{\text{H}}=\frac{1}{2\pi}\left[r_{\text{h}}\left(1+\frac{r^2_{\text{h}}}{l^2}\right)\left(\frac{1}{2\rh^2}+ \frac{1}{\rh^2+a^2}\right)-\frac{1}{\rh}\right],
\end{equation}
and the entropy is
\begin{equation}
  S=\frac{\pi(\rh^2+a^2)}{\Xi}.
\end{equation}
In the extended phase space, the negative cosmological constant is interpreted as pressure~\cite{KaRT09}
\begin{equation}
  P=\frac{3}{8\pi}\frac{1}{l^2},
\end{equation}
with the conjugate thermodynamic variable
\begin{equation}
  V=\frac{4\pi}{3}\frac{\rh(\rh^2+a^2)}{\Xi}\left(1+\frac{a^2}{2\rh^2}\frac{1+\frac{\rh^2}{l^2}}{\Xi}\right),
\end{equation}
 and the specific volume
\begin{equation}\label{SpV}
  v=2\left(\frac{3V}{4\pi}\right)^{\frac{1}{3}}=2 \left[ \frac{\rh(\rh^2+a^2)}{\Xi}\left(1+\frac{a^2}{2\rh^2}\frac{1+\frac{\rh^2}{l^2}}{\Xi}\right) \right]^{\frac{1}{3}}.
\end{equation} 
From the above thermodynamic variables, it is clear that they satisfy the first law of black hole thermodynamics and the Smarr relation
\begin{gather}
  dM = T_{\rm H}dS+\Omega_{\text{H}}dJ+VdP, \label{1srlaw} \\
  M = 2T_{\text{H}}S+2\Omega_{\text{H}}J-2PV.
\end{gather}

Having the thermodynamic quantities and the first law, we can investigate the critical point and coexistence curve for black hole phase transition through the Maxwell's equal-area law.

\subsection{Exact coexistence curve and order parameter}\label{Sec:coexi}

As it is well known, critical phenomena and coexistence curves are extremely important in the study of phase transition. Thermodynamics along the coexistence curve usually exhibits exciting phenomena. For later convenience of the discussion for the kinetic properties along the coexistence curve, we briefly review the coexistence curve and discuss the order parameter in this part.

Usually, we use Maxwell's equal-area law to obtain the coexistence curve for black hole phase transition~\cite{SpSm13,ZhWe19,ZhLW20}. However, due to the calculation difficulty for the coexistence curve for the phase transition of the Kerr-AdS black hole, it is very hard to solve analytically the equations for the coexistence curve and critical point in practical.

By expanding all thermodynamic quantities to $\mathcal{O}(J^2)$ in the small-$J$ limit, Gunasekaran et al. obtained an analytical result of critical point for slowly rotating black holes~\cite{GuKM12}. In Refs.~\cite{WeLi14,WeCL16}, some of the current authors classified the thermodynamic quantities into universal and characteristic parameters, and argued that the critical point can be interpreted as the relation between the universal and characteristic parameters. According to the classification, the only characteristic thermodynamic quantity for the Kerr-AdS black hole is the angular momentum $J$. From dimensional analysis, the critical point should have the form
\begin{align}\label{critForm}
P_{\text{c}}&=\frac{\alpha_1}{J}, &\qquad T_{\text{c}}&=\frac{\alpha_2}{\sqrt{J}},  &\qquad S_{\text{c}}=\alpha_3 J, \notag \\
V_{\text{c}}&=\alpha_4 J^{\frac{3}{2}}, &\qquad v_{\text{c}}&=\alpha_5 \sqrt{J}.
\end{align}
The dimensionless parameters $\alpha_1$--$\alpha_5$ can be determined by the $S$--$T_{\rm H}$ oscillatory curve.

Using the $S$--$T_{\rm H}$ oscillatory curve, the critical point can be determined by
\begin{gather}
  \left( \frac{\partial T_{\rm H}}{\partial S}\right)_{J,P} = 0,\\
  \left( \frac{\partial^2 T_{\rm H}}{\partial S^2}\right)_{J,P} = 0.
\end{gather}

Notwithstanding in a complicated form, the equations for the parameters $\alpha_1$--$\alpha_5$ can be solved analytically. The approximate values for the critical point for all values of the angular momentum $J$ are given by~\cite{WeCL16}
\begin{align}\label{CritPoint}
  P_{\text{c}}&\approx\frac{0.002857}{J},  &\qquad    T_{\text{c}}&\approx\frac{0.041749}{\sqrt{J}},  \notag \\
  V_{\text{c}}&\approx 115.796503 J^ {\frac{3}{2}}, &\qquad  v_{\text{c}}&\approx 6.047357 \sqrt{J}.
\end{align}

According to the swallow tail behavior of the Gibbs free energy $G$ and the Maxwell's equal-area law, the fitting formula for the coexistence curve can be obtained~\cite{WeCL16}
\begin{multline}
  \tilde{P} = 0.718781
   \tilde{T}^2+0.188586 \tilde{T}^3+0.061488 \tilde{T}^4 \\
   + 0.022704 \tilde{T}^5 +0.002340 \tilde{T}^6+0.010547 \tilde{T}^7\\
  -0.008649 \tilde{T}^8
   + 0.005919
   \tilde{T}^9-0.001717 \tilde{T}^{10},  \label{coexistenceCurve}
\end{multline}
where the reduced parameters are defined as
\begin{align}
  \tilde{P}=\frac{P}{P_{\text{c}}}, &\qquad \tilde{T}=\frac{T_{\rm H}}{T_{\text{c}}}.\label{reducedPara}
\end{align}
It is intriguing that the coexistence curve in the reduced parameter space is angular momentum independent.

\begin{figure}
  \centering
  \includegraphics[width=2.8in]{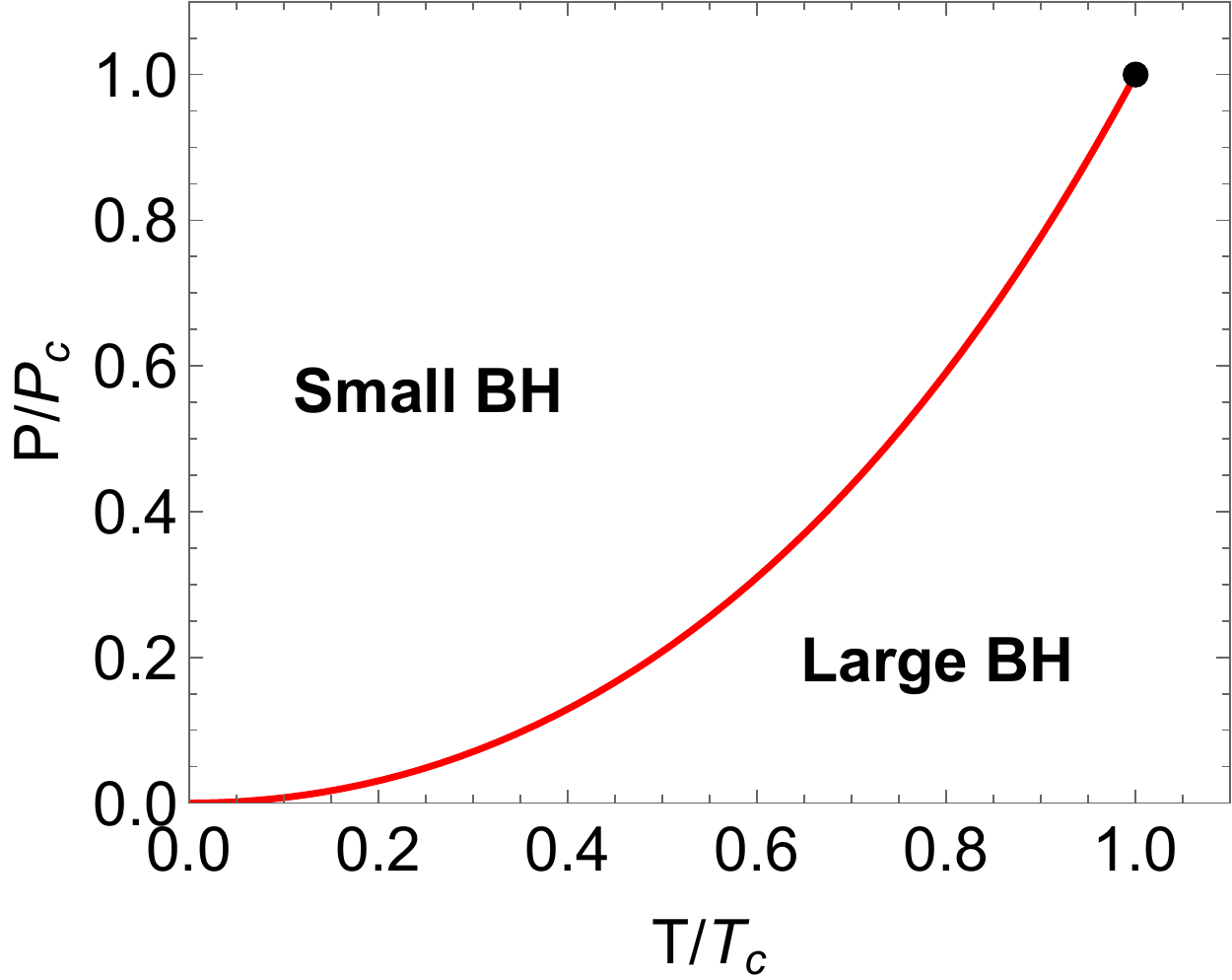}
  \caption{The phase diagram for the Kerr-AdS black hole in the reduced parameter space. The red line denotes the small-large black hole coexistence curve. The black dot located at $(1,1)$ is the critical point. Above the coexistence curve is the small black hole phase. Below the coexistence curve is the large black hole phase.}\label{Fig: coexistence}
\end{figure}

The phase diagram in the reduced parameter space is depicted in Fig.~\ref{Fig: coexistence}.
The red line in Fig.~\ref{Fig: coexistence} shows the coexistence curve for small-large black hole phase transition. The coexistence line terminates at the critical point, beyond which there is no small-large black hole phase transition. We denote the black hole event horizon radius at the critical point by $r_{\text{c}}$.

The order parameter changes discontinuously below the critical point but continuously at the critical point. As we can see in Fig.~\ref{Fig: Gibbs} for the swallow tail behavior of the Gibbs free energy discussed in Sec.~\ref{Sec:GibbsFE}, a sudden change of the black hole horizon radius difference $r_{\text{h}}-r_{\text{c}}$ occurs for the small-large black hole phase transition below the critical point, and the change of the horizon radius difference $r_{\text{h}}-r_{\text{c}}$ becomes zero at the critical point. Hence, the horizon radius difference $r_{\text{h}}-r_{\text{c}}$ can be regarded as the order parameter for the small-large black hole phase transition. Since the horizon radius $r_{\text{h}}$ is just a parallel transport of the horizon radius difference $r_{\text{h}}-r_{\text{c}}$, we can use the horizon radius $r_{\text{h}}$ as the order parameter for convenience. In the following discussion of the black hole phase transition, we regard the black hole horizon radius $r_{\text{h}}$ as the order parameter.

\section{The free energy landscape}\label{Sec:lands}

\subsection{Gibbs free energy}\label{Sec:GibbsFE}

Besides the Maxwell's equal area law, the Gibbs free energy is also a powerful tool to investigate black hole phase transition. {The Gibbs free energy exhibits a swallow tail behavior and its first-order derivative is discontinuous at the first-order phase transition point.  For a second-order phase transition, the Gibbs free energy and its first-order derivative are continuous but not smooth.}

The Gibbs free energy for the Kerr-AdS black hole is
\begin{multline}
 G =M-T_{\rm H}S \\
              =\frac{(\rh^2+a^2)(1+\frac{\rh^2}{l^2})}{2\Xi^2\rh} -\frac{\rh}{4\Xi} \left(1+\frac{a^2}{l^2}+\frac{3\rh^2}{l^2}-\frac{a^2}{\rh^2} \right).\label{Gibbs}
\end{multline}
We can express the Gibbs free energy $G$ and temperature $T_{\rm H}$ in terms of the pressure $P$, entropy $S$, and angular momentum $J$ \cite{WeCL16}
\begin{gather}
  G = \frac{12\pi^2J^2(16PS+9)-64P^2S^4+9S^2}{12\sqrt{\pi S}\sqrt{8PS+3}\sqrt{12\pi^2J^2+S^2(8PS+3)}} ,\\
  T_{\rm H} = \frac{S^2(64P^2S^2+32PS+3)-12\pi^2J^2}{4\sqrt{\pi}S^{\frac{3}{2}}\sqrt{8PS+3}\sqrt{12\pi^2J^2+S^2(8PS+3)}}.
\end{gather}

\begin{figure}
  \centering
  \includegraphics[width=3.1in]{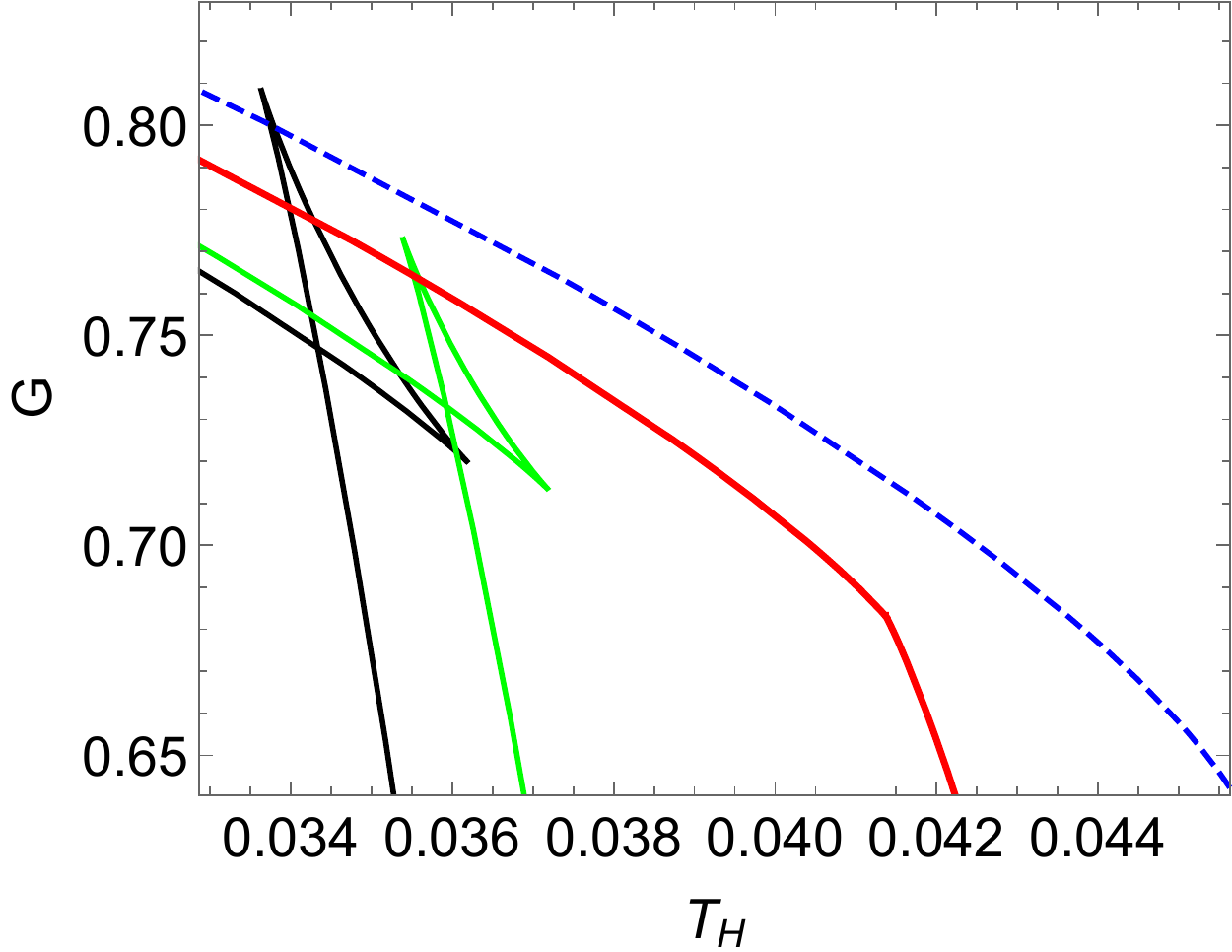}
  \caption{The Gibbs free energy for the Kerr-AdS black hole with fixed angular momentum $J=1$ for various values of the pressure $P$. The pressure decreases from right to left. The blue dashed line corresponds to $P=1.25 P_{\rm c}$, the red thick solid line to $P=P_{\rm c}$. The remaining solid lines correspond to $P= 0.72P_{\rm c}$ and $P= 0.64P_{\rm c}$, respectively. For pressure lower than the critical pressure $P_{\rm c}$, the Gibbs free energy shows the swallow tail behavior and the system displays a first-order phase transition. {The horizon radius increases from left to right along the curve, and a sudden change of the horizon radius occurs in the small-large black hole phase transition point.}}\label{Fig: Gibbs}
\end{figure}

We plot the Gibbs free energy as a function of the black hole temperature $T_{\rm H}$ with fixed angular momentum $J$ and various thermodynamic pressure $P$ parametrically in Fig.~\ref{Fig: Gibbs}.

The Gibbs free energy $G$ demonstrates the characteristic swallow tail behavior when the pressure $P$ is lower than the critical pressure $P_{\text{c}}$. The swallow tail behavior indicates a small-large black hole phase transition. While the swallow tail behavior disappears and there is no phase transition when the pressure is higher than the critical pressure.

The thermodynamic stability of a black hole is determined by its heat capacity. For positive heat capacity, the black hole is locally thermodynamically stable; while negative heat capacity indicates thermodynamic instability. For a canonical ensemble in the extended phase space, we consider the heat capacity
\begin{equation}\label{heatcapacity}
  C_{P,J}=T_{\rm H}\left(\frac{\partial S}{\partial T_{\rm H}}\right)_{P,J}.
\end{equation}
In Fig.~\ref{Fig: Gibbs}, the intermediate black hole has negative heat capacity and so it is thermodynamically unstable, while the small and large black holes are locally thermodynamically stable.

A canonical ensemble is defined as the ensemble with fixed temperature. We consider the  canonical ensemble at a specific temperature $T_{\rm E}$ composed of a series of black holes with various black hole horizon radius. It is clear that, only when the Hawking temperature of the black hole is the same as the temperature of the ensemble $T_{\rm E}$, can the black hole be in equilibrium.

To make progress and understand the physical process of phase transition and thermodynamic stability, we exploit the off shell Gibbs free energy.

The off shell Gibbs free energy is defined by replacing the black hole temperature $T_{\rm H} $ in the formula of Gibbs free energy with the temperature of the ensemble $T_{\rm E}$,
\begin{equation}\label{DGibbs}
  G_{\text{L}}=M-T_{\text{E}}S.
\end{equation}
For the Kerr-AdS black hole, the off shell Gibbs free energy is
\begin{equation}\label{KGibbs}
  G_{\text{L}}=\frac{1}{2\rh}\frac{(\rh^2+a^2)(1+\frac{\rh^2}{l^2})}{\Xi^2} -T_{\text{E}}\frac{\pi (\rh^2+a^2)}{\Xi}.
\end{equation}

The black hole temperature $T_{\text{H}}$ is a function of black hole horizon radius $\rh$, angular momentum $J$, and pressure $P$. However, the temperature of the ensemble $T_{\rm E} $ has nothing to do with these parameters, and it is an independent thermodynamic variable. Hence, the off shell Gibbs free energy can be regarded as a function of the black hole event horizon radius $\rh$, angular momentum $J$, pressure $P$, and temperature of the ensemble $T_{\text{E}}$,
\begin{equation}
  G_{\text{L}}=G_{\text{L}}(\rh, J, P, T_{\text{E}}).
\end{equation}

The name ``off shell'' just means not all points on the curve of $\GL$ correspond to a physical situation, since the corresponding black holes do not have the correct temperature and they are not equilibrium configurations~\cite{AnLe20}.  But the thermodynamic quantities still satisfy the black hole thermodynamic relation. In some literature the off shell Gibbs free energy is also called the generalized Gibbs free energy~\cite{York86}.

\begin{figure}
  \centering
  \includegraphics[width=3.2in]{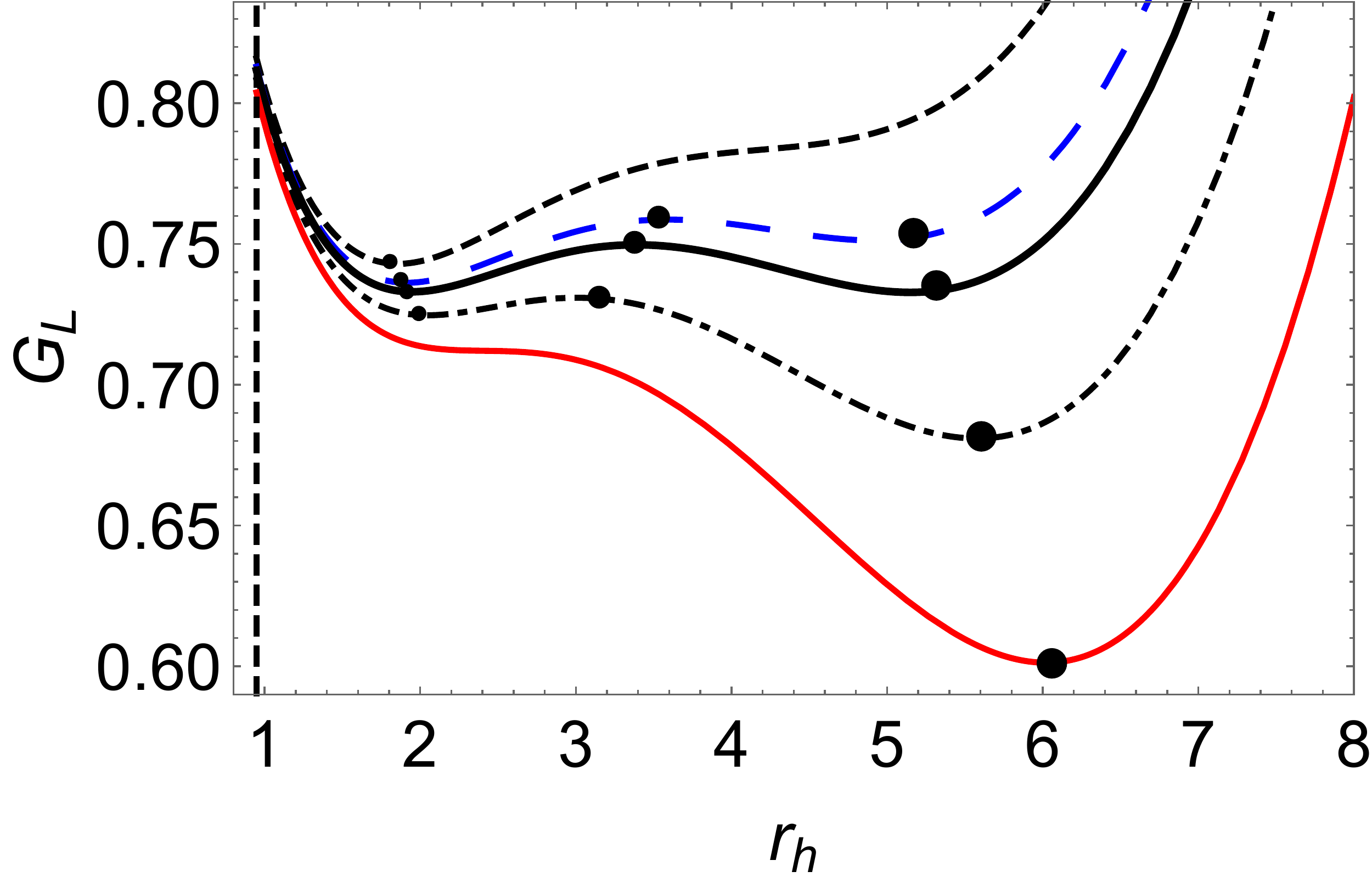}
  \caption{The off shell Gibbs free energy for the Kerr-AdS black hole with various temperature $T_{\rm E}$. Here we have set the angular momentum $J=1$ and the thermodynamic pressure $P=0.72P_{\rm c}$. Physical black holes are located at the extremal points of the off shell Gibbs free energy. The vertical dashed line is the lower bound for the black hole event horizon as discussed in Sec.~\ref{Sec:lowbound}.}\label{offG}
\end{figure}

As proved in the previous work~\cite{WeLW20}, a physical black hole and its stability are related to the extremal point of the off shell Gibbs free energy. Using the first law of black hole thermodynamics~\eqref{1srlaw} and the off shell Gibbs free energy~\eqref{KGibbs}, we have
\begin{equation}
\begin{split}
    \left(\frac{\partial \GL}{\partial \rh}\right)_{P,J,T_{\text{E}}} &=  \left(\frac{\partial\GL}{\partial S}\right)_{P,J,T_{\text{E}}}\left(\frac{\partial S}{\partial \rh}\right)_{P,J}  \\
   &=\left[\left(\frac{\partial M}{\partial S}\right)_{P,J}-T_{\text{E}}\right]\left(\frac{\partial S}{\partial \rh}\right)_{P,J}  \\
  & = \left(T_{\text{H}}-T_{\text{E}}\right)\left(\frac{\partial S}{\partial \rh}\right)_{P,J}.
\end{split}
\end{equation}
It is clear that a physical black hole corresponds to an extremal point of the off shell Gibbs free energy as indicated in Fig.~\ref{offG}.

The thermodynamic stability of the black hole can be examined from the second-order derivative of the off shell Gibbs free energy. In the extremal point of the off shell Gibbs free energy, the second-order derivative of the off shell Gibbs free energy is
\begin{equation}
\begin{split}
  \left.\left(\frac{\partial^2 \GL}{\partial \rh^2}\right)_{P,J,T_{\text{E}}}\right|_{T_{\rm E}=T_{\rm H}} &=\left( \frac{\partial T_{\text{H}}}{\partial\rh}\right)_{P,J}\left(\frac{\partial S}{\partial \rh}\right)_{P,J} \\
     & =\left( \frac{\partial T_{\text{H}}}{\partial S}\right)_{P,J}\left(\frac{\partial S}{\partial \rh}\right)_{P,J}^2  \\
     & =\frac{T_{\text{H}}}{C_{P,J}} \left(\frac{\partial S}{\partial \rh}\right)_{P,J}^2.
\end{split}
\end{equation}

It is clear that each black hole with a maximum value of the off shell Gibbs free energy has negative heat capacity and is thermodynamically unstable; while the heat capacity of the black hole at minimum point of the off shell Gibbs free energy is positive, which indicates that such black hole is thermodynamically stable.

Each extremal point of the off shell Gibbs free energy corresponds to a physical black hole. How about the behavior of the off shell Gibbs free energy for vanishing black hole event horizon radius? We explore this issue in the next subsection.

\subsection{The lower bound for the order parameter}\label{Sec:lowbound}

Previous research for the charged AdS black hole~\cite{LiZW20} argues that the off shell Gibbs free energy can be regarded as a function of the black hole horizon radius $\rh$, charge $Q$, thermodynamic pressure $P$, and temperature of the ensemble $T_{\text{E}}$, and the black hole horizon radius can be any positive value. In particular, the work argues that the off shell Gibbs free energy is divergent for vanishing black hole horizon radius~\cite{LiZW20}.

However, these contradict the physical intuition that vanishing black hole event horizon radius corresponds to a thermal AdS space. The enthalpy and entropy of the thermal AdS space are zero. Hence, the off shell Gibbs free energy should be zero. In addition, the thermal AdS space is also a solution of Einstein field equations, it should be a point in the off shell Gibbs free energy.

The reason for the contradiction is that the black hole thermodynamic quantities and the relation between them were used but the fact that the metric should describe a black hole was neglected. When one sets the black hole charge $Q=1$, the black hole event horizon can never approach zero. This means that there is a lower bound for the event horizon for the Kerr-Newman-AdS family black holes. The lower bound is a function of the charge $Q$, angular momentum $J$, and pressure $P$, i. e.
\begin{equation}
  \rh\geq \re(Q,J,P).
\end{equation}
For the Kerr-Newman-AdS family black holes, the lower bound corresponds to the event horizon of the extremal black hole.

For a given angular momentum $J$ and pressure $P$, the lower bound for the order parameter for small-large black hole phase transition of the Kerr-AdS black hole corresponds to the extremal Kerr-AdS black hole.

\begin{figure}
  \centering
  \includegraphics[width=2.8in]{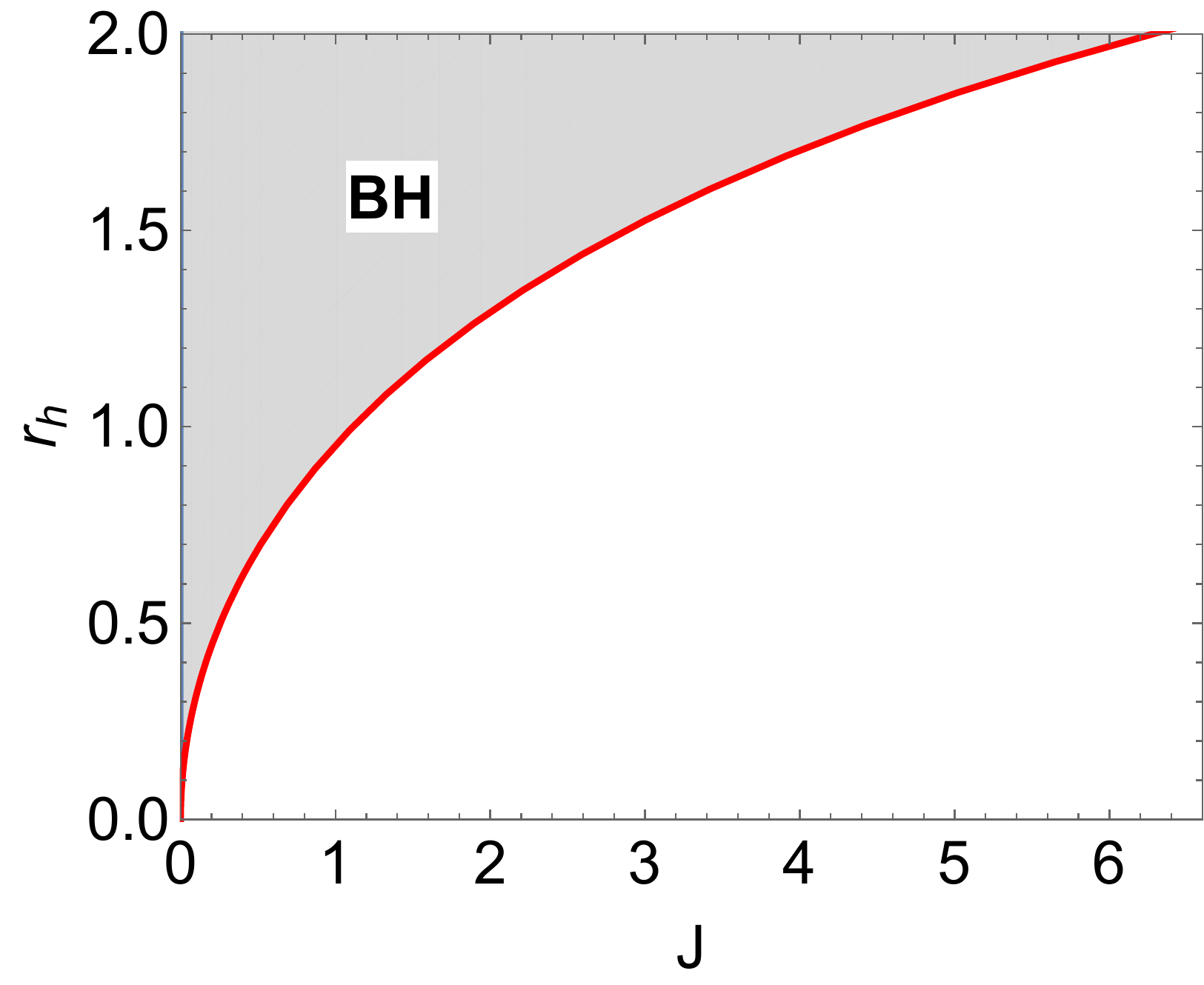}
  \caption{The lower bound of the event horizon radius for the Kerr-AdS black hole with pressure $P=0.0018$. The red thick line is the lower bound for the black hole event horizon. Above the red thick line is the black hole region which is the allowed region for values of the black hole event horizon.}\label{fig:Lbound}
\end{figure}

The event horizon of the extremal Kerr-AdS black hole $\re$ is determined by
\begin{equation}\label{extremrh}
  \Delta=\left(\re^2+a^2\right)\left(1+\frac{\re^2}{l^2}\right)-2m\re=0.
\end{equation}
The inner and outer horizons of the extremal Kerr-AdS black hole coincide. Hence, the event horizon is degenerate. Then, we have
\begin{equation}\label{DeriveDelta}
  \Delta'(\re)=\frac{4}{l^2}\re^3+2\left(1+\frac{a^2}{l^2}\right)\re-2m=0.
\end{equation}
From Eqs.~\eqref{extremrh} and~\eqref{DeriveDelta}, we have
\begin{equation}\label{reEq}
  \frac{3}{l^2}\re^4+\left(1+\frac{a^2}{l^2}\right)\re^2-a^2=0.
\end{equation}
Then, the event horizon of the extremal Kerr-AdS black hole is
\begin{equation}\label{horizonre}
  \re=\left[\frac{-\left(l^2+a^2\right) +\sqrt{\left(l^2+a^2\right)^2+12a^2l^2}}{6}\right]^{\frac{1}{2}}.
\end{equation}
Hence, the event horizon of the Kerr-AdS black hole satisfies
\begin{equation}
\begin{split}
    \rh & \geq  \re(J,P)\\
     &  = \left[\frac{\sqrt{(3+8\pi Pa^2)^2+288\pi P a^2}-(3+8\pi Pa^2)}{48\pi P}\right]^{\frac{1}{2}}.
\end{split}
\end{equation}
The allowed region for the black hole event horizon is plotted in Fig~\ref{fig:Lbound}.

It is clear from Fig.~\ref{fig:Lbound} that the horizon radius of the Kerr-AdS black hole can never be zero for nonvanishing angular momentum.
For a fixed angular momentum, the larger the black hole mass $M$, the bigger the black hole. From Fig.~\ref{fig:Lbound} and Eq.~\eqref{reEq}, the angular momentum is zero for a vanishing black hole horizon radius. Furthermore, the mass of the black hole should also be zero for a vanishing black hole horizon radius. Thus the space is indeed an AdS space for a vanishing horizon radius.

\begin{figure}
  \centering
  \includegraphics[width=3.1in]{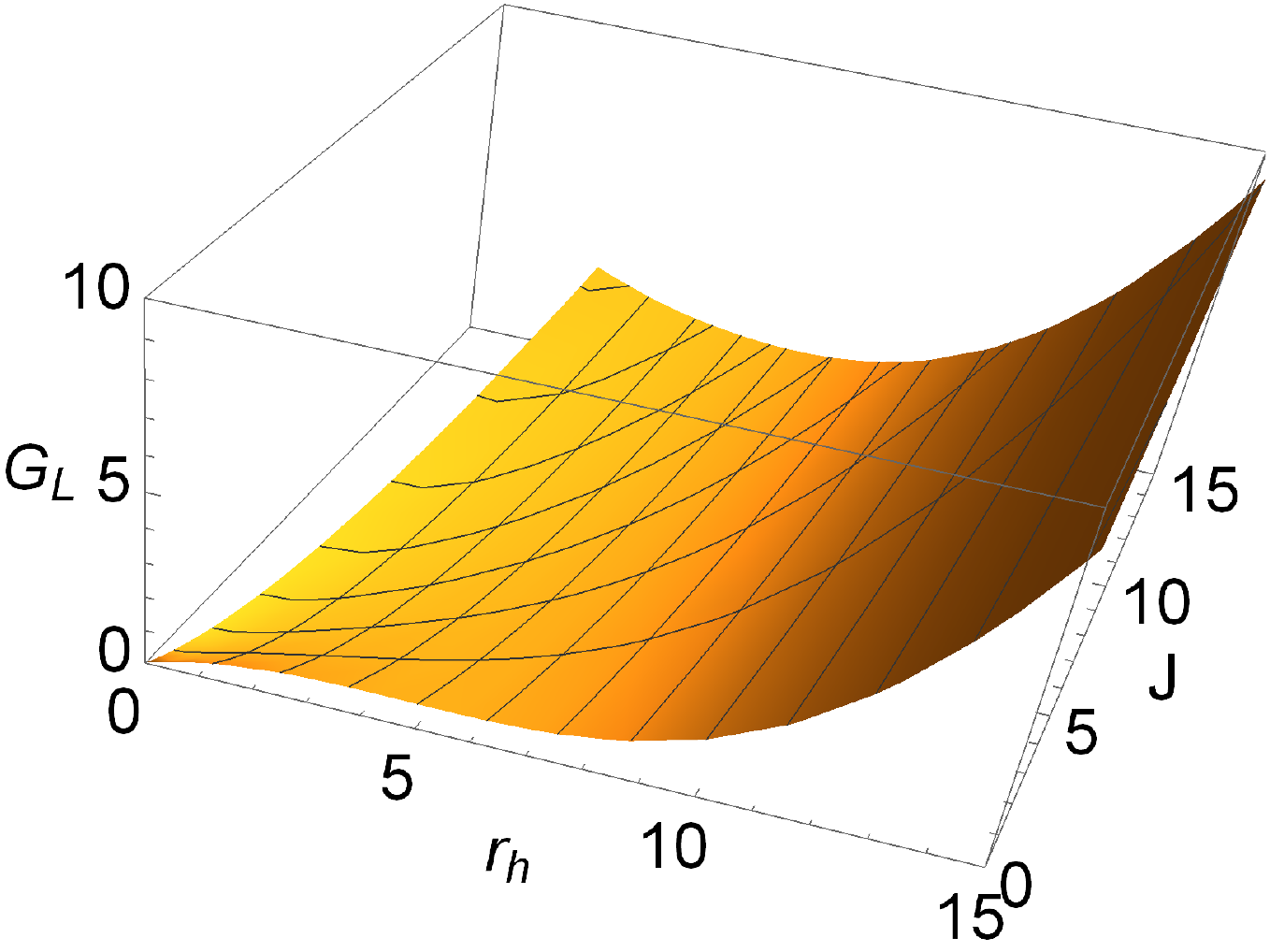}
  \caption{The off shell Gibbs free energy $\GL$ as a function of the black hole event horizon $\rh$ and angular momentum $J$ for the Kerr-AdS black hole. We have set the pressure $P=0.0018$ and temperature of the ensemble $T_{\rm E}=0.03433$. The off shell Gibbs free energy is zero for vanishing black hole horizon radius.}\label{fig:3DGibbs}
\end{figure}

Now, we can investigate the behavior of the off shell Gibbs free energy near the point of vanishing horizon radius. We plot the 3D figure for the off shell Gibbs free energy $\GL(\rh,J)$ with fixed pressure $P$ and temperature of the ensemble $T_{\text{E}}$ parametrically according to the following relations
\begin{align}
  \rh &= \re+r'_{\text{h}}, \\
  J &= \frac{3a}{2\rh}\frac{(\rh^2+a^2)(3+8\pi P\rh^2)}{(3+8\pi Pa^2)^2}, \\
  \GL &= \frac{3}{2\rh}\frac{(\rh^2+a^2)(3+8\pi P\rh^2)}{(3-8\pi Pa^2)^2} -\frac{3\pi (\rh^2+a^2)}{3-8\pi Pa^2}T_{\text{E}},
\end{align}
where $\rh'$ is zero or any positive parameter. From Fig.~\ref{fig:3DGibbs}, it is clear that the off shell Gibbs free energy approaches zero for vanishing black hole event horizon radius, and it corresponds to the thermal AdS space.

The extremal points of the off shell Gibbs free energy $\GL(\rh)$ correspond to the physical black holes, since these points satisfy the laws of black hole thermodynamics. However, the thermal AdS space does not satisfy the laws of black hole thermodynamics. Hence it is the origin instead of extremal point of the off shell Gibbs free energy.

It is also clear from Fig. \ref{fig:3DGibbs} that for fixed nonzero angular momentum, there is no Hawking-Page phase transition though the extremal points of the off shell Gibbs free energy might equal to zero, since the thermal AdS space is at the origin with vanishing angular momentum.

\subsection{The landscape}\label{Sec:thelandscape}

On the free energy landscape, only the extremal point of the off shell Gibbs free energy describes the physical black hole, and the Hawking temperature of each black hole $T_{\text{H}}$ is the same as the temperature of the ensemble $T_{\text{E}}$. For fixed angular momentum $J$, pressure $P$, and temperature of the ensemble $T_{\text{E}}$, we can investigate the off shell Gibbs free energy $\GL$ as a function of black hole horizon radius $\rh$, which is the order parameter in small-large black hole phase transition for the Kerr-AdS black hole. From the free energy landscape perspective, black hole phase transition is regarded as the stochastic fluctuation of the order parameter. The off shell Gibbs free energy $\GL(\rh)$ plays the role of a driving force for the phase transition. Hence, the behavior of the off shell Gibbs free energy is pivotal in the discussion of kinetic process of phase transition. In this subsection, we exploit the behavior of the off shell Gibbs free energy.

Without loss of generality, we take the angular momentum $J=1$, the thermodynamic pressure $P=0.64P_{\text{c}}$. As discussed in the previous subsection, there is a lower bound on black hole event horizon as shown in Fig.~\ref{fig:Lbound}. In the present case, the lower bound for the black hole event horizon is approximately $r_{\rm ex}=0.953$.

\begin{figure*}[!htb]
\begin{center}
  \subfigure[]{\includegraphics[width=3.2in,height=2.2in]{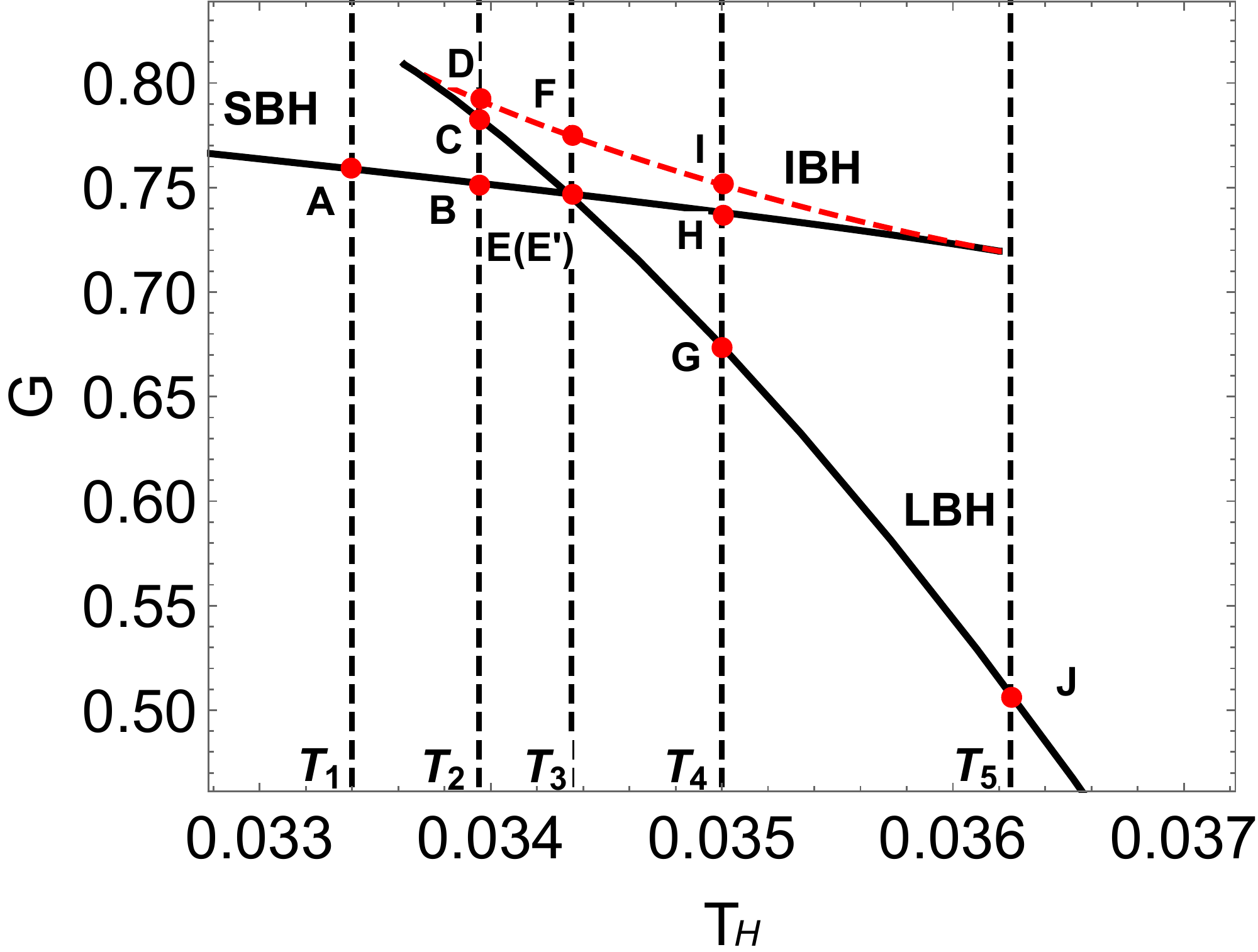}\label{Fig:off-shella}}
\subfigure[]{\includegraphics[width=3.2in]{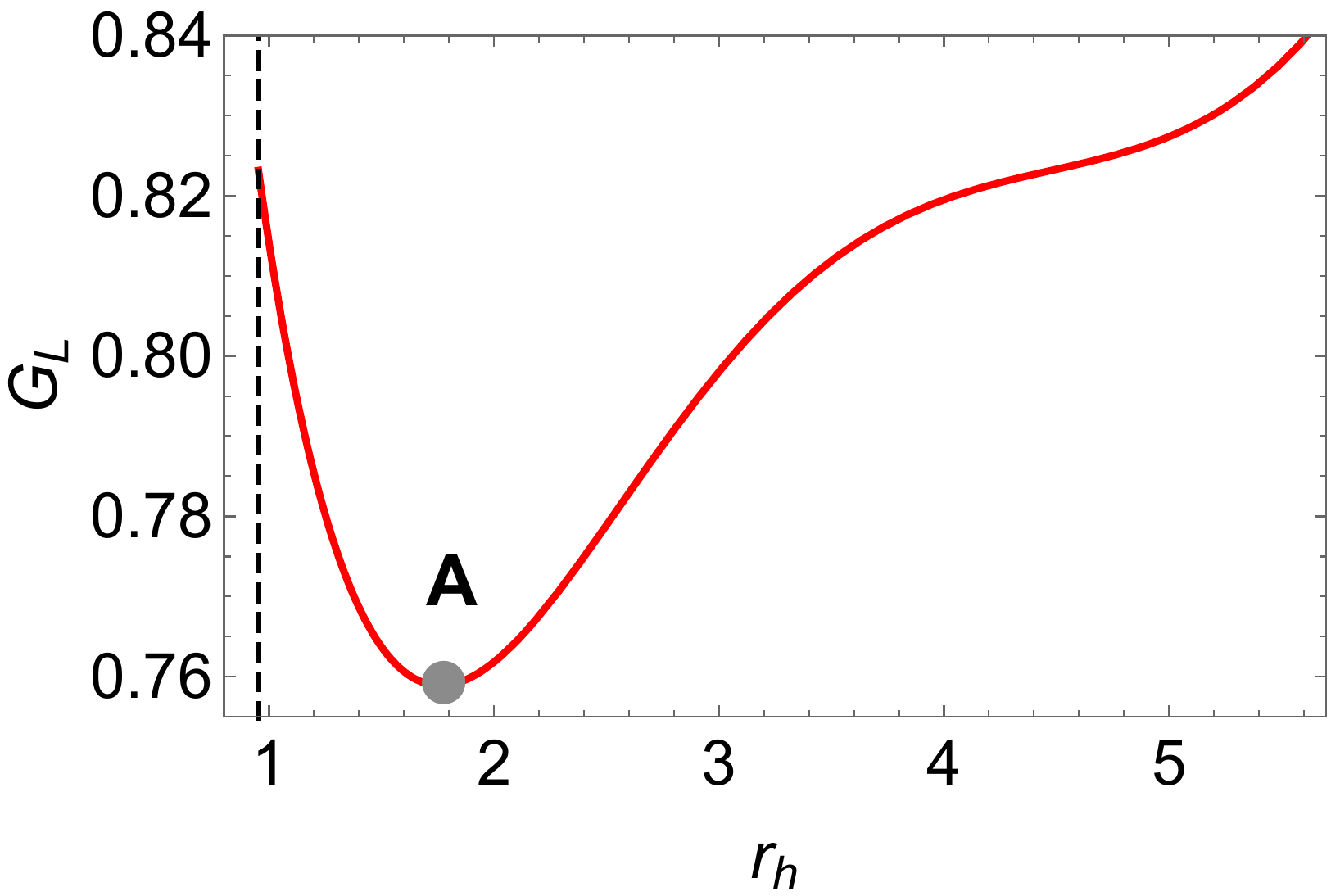}\label{Fig:off-shellb}}
\subfigure[]{\includegraphics[width=3.2in]{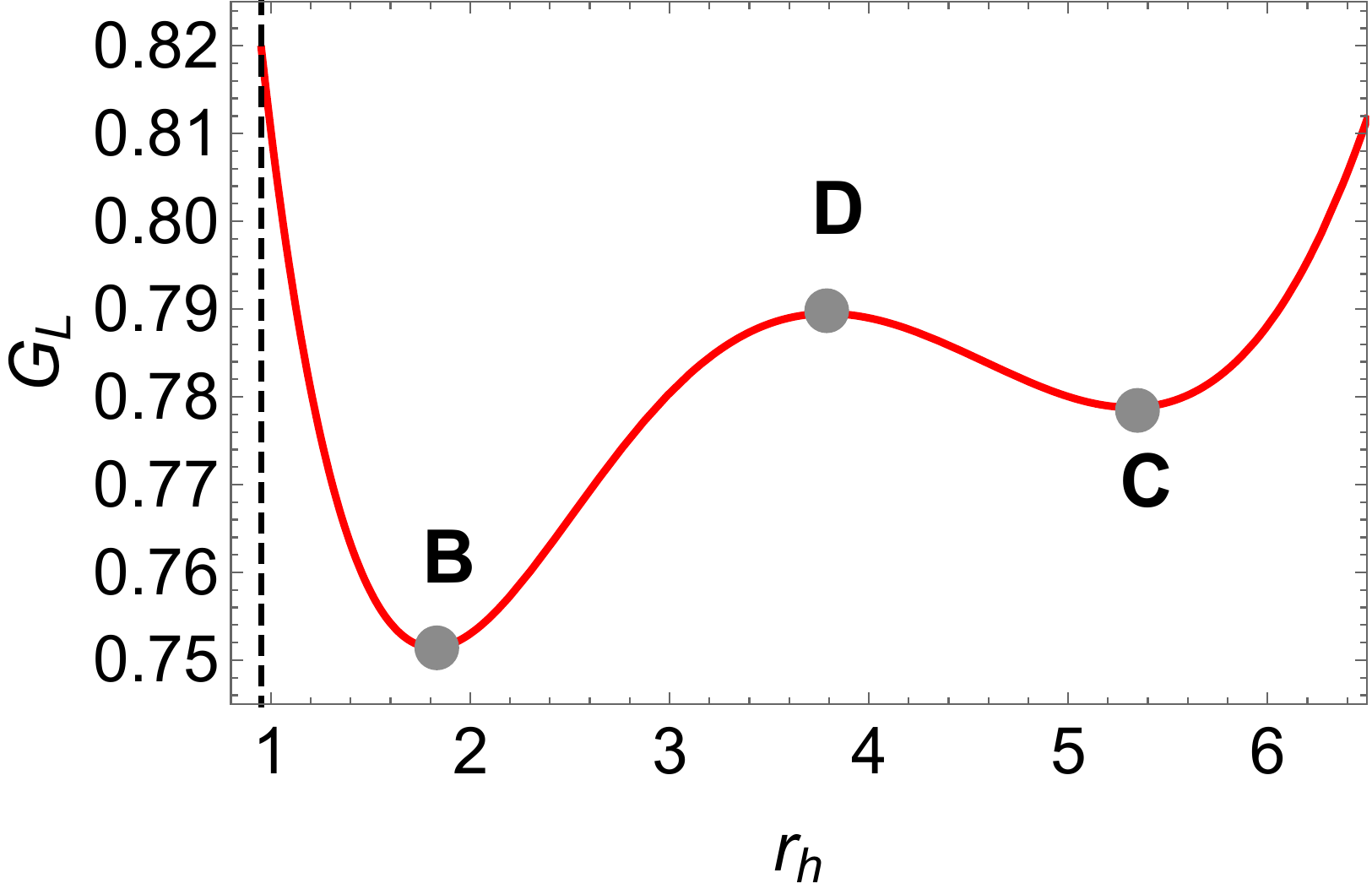}\label{Fig:off-shellc}}
\subfigure[]{\includegraphics[width=3.2in]{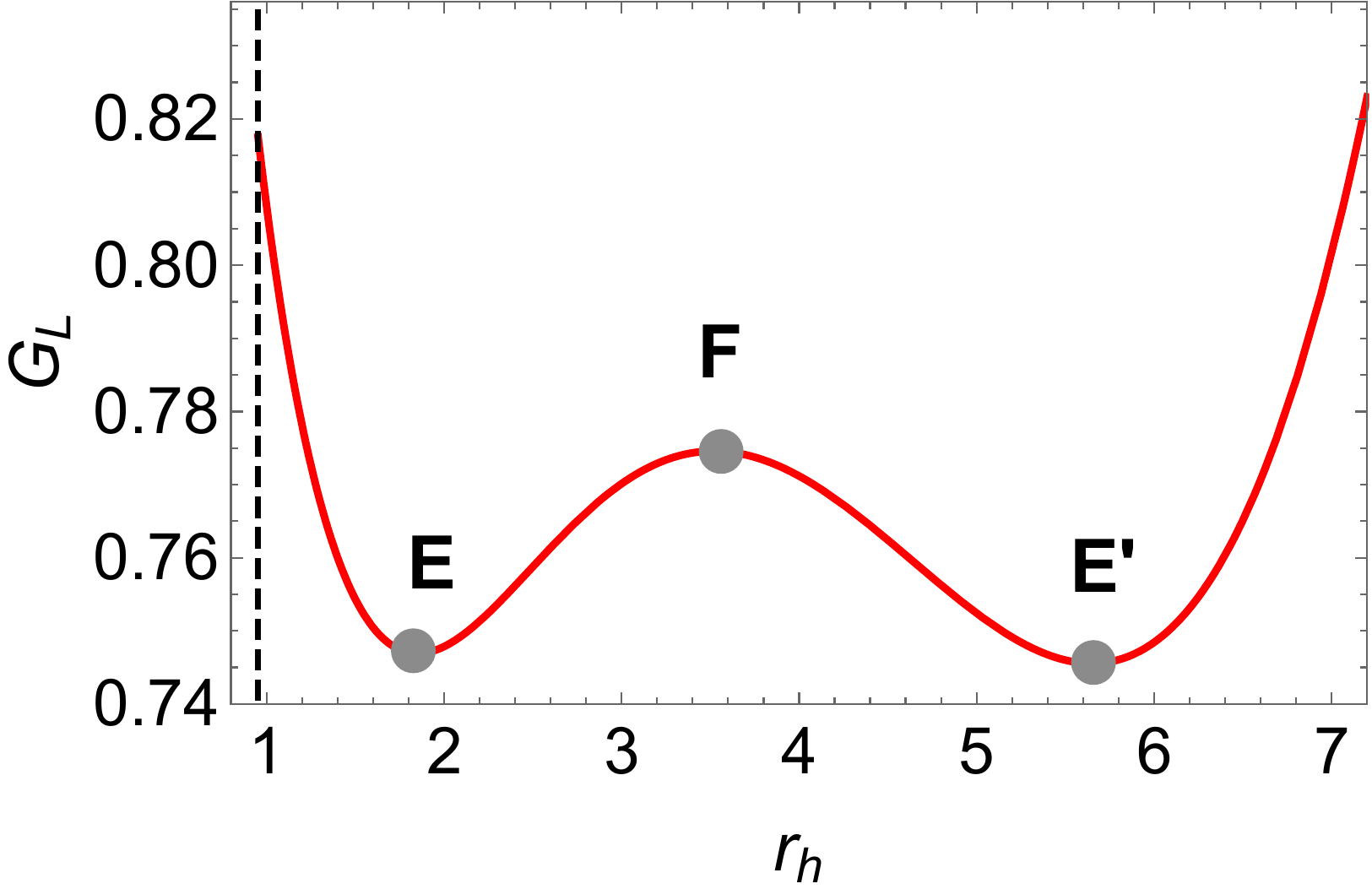}\label{Fig:off-shelld}}
\subfigure[]{\includegraphics[width=3.2in]{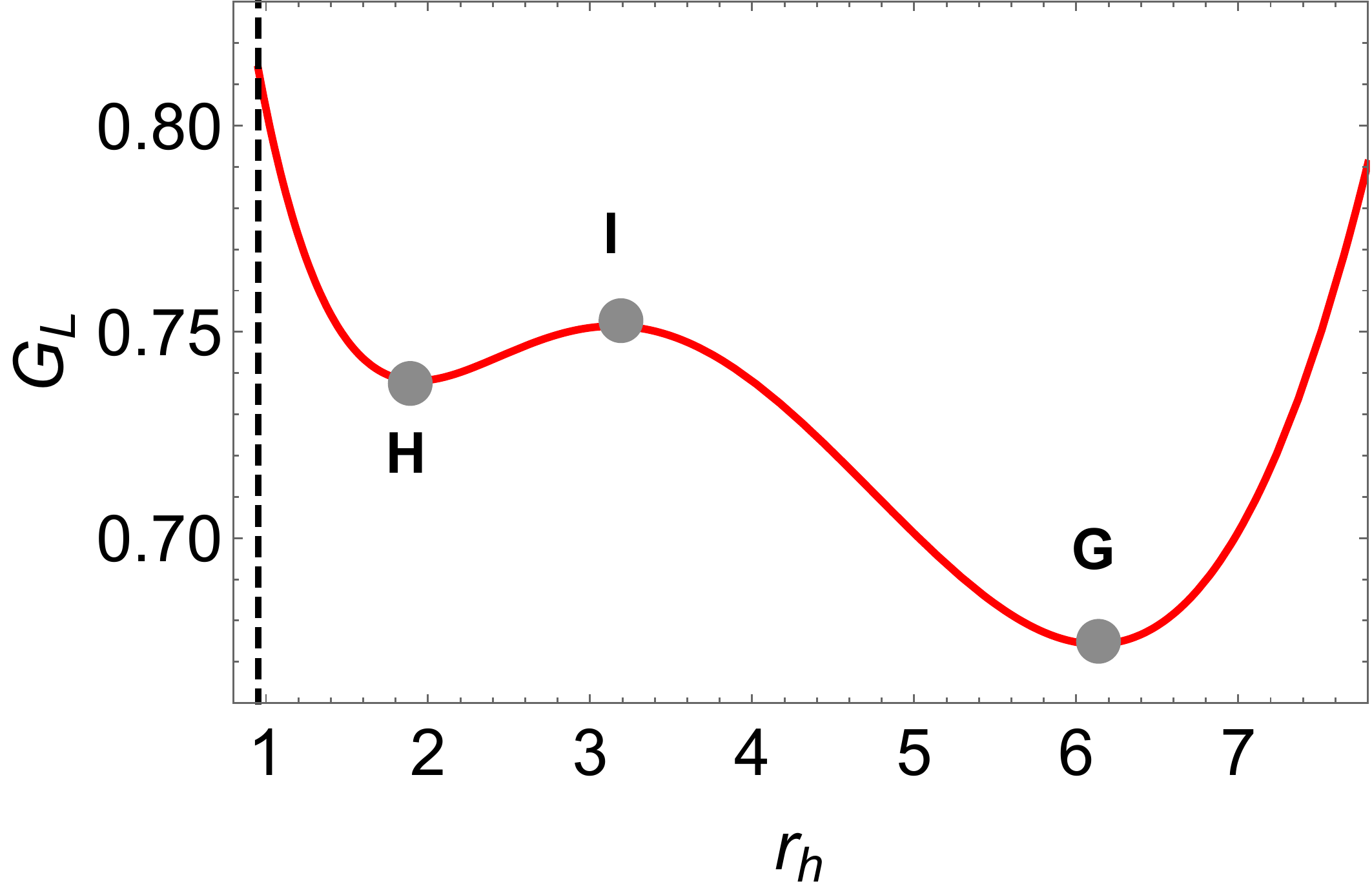}\label{Fig:off-shelle}}
\subfigure[]{\includegraphics[width=3.2in]{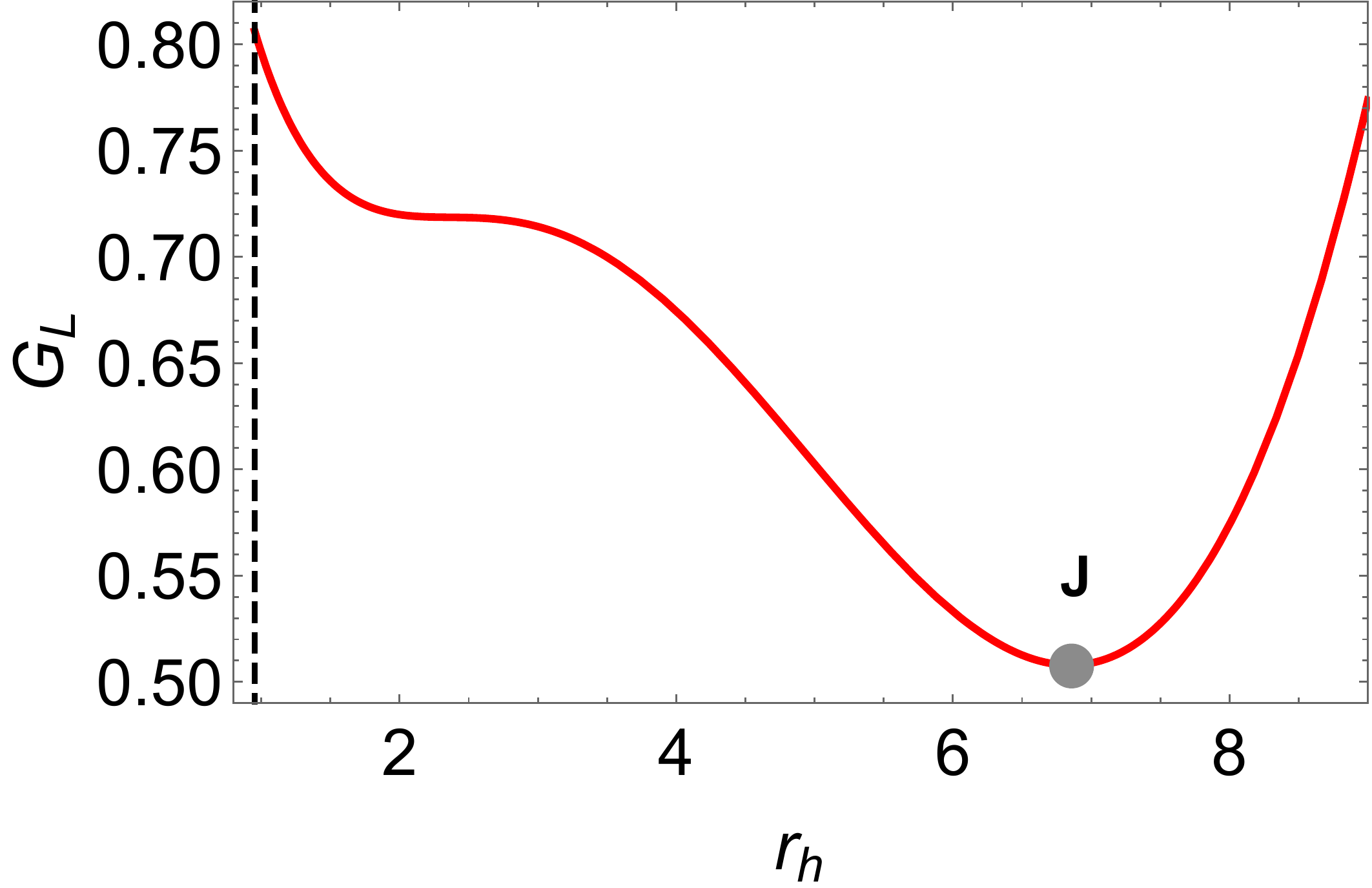}\label{Fig:off-shellf}}
  \caption{The behavior of the Gibbs free energy and off shell Gibbs free energy. Where we set the angular momentum of the black hole $J=1$ and the thermodynamic pressure $P=0.64P_{\rm c}$. (a)~The Gibbs free energy $G$ vs the black hole temperature $T_{\rm H}$.  The temperatures $T_1- T_5$ are equal to $0.8080T_{\rm c}, 0.8212T_{\rm c}, 0.8306T_{\rm c}, 0.8467T_{\rm c}$ and $0.8769T_{\rm c}$, respectively. (b)~The off shell Gibbs free energy $\GL$ vs the black hole event horizon $\rh$ with the temperature of the ensemble $T_{\rm E}=T_1$. (c)~$\GL$ vs  $\rh$ with $T_{\rm E}=T_2$. (d)~$\GL$ vs  $\rh$ with $T_{\rm E}=T_3$. (e)~$\GL$ vs  $\rh$ with $T_{\rm E}=T_4$. (f)~$\GL$ vs  $\rh$ with $T_{\rm E}=T_5$.}\label{Fig:off-shell}
 \end{center}
\end{figure*}

To make the relationship between the Gibbs free energy $G$ and the off shell Gibbs free energy $\GL$ clearer, we display the figures for the Gibbs free energy $G(T_{\rm H})$ and off shell Gibbs free energy $\GL(\rh)$ with various temperatures of the ensemble together in Fig.~\ref{Fig:off-shell}.

The behavior of the Gibbs free energy $G$ as a function of black hole temperature $T_{\rm H}$ is shown in Fig.~\ref{Fig:off-shella}. The swallow tail behavior demonstrates the small-large black hole phase transition. The red dashed line corresponds to intermediate black holes, which are thermodynamically unstable as discussed in Sec.~\ref{Sec:GibbsFE} from the heat capacity viewpoint; while the small and large black holes are locally thermodynamically stable. Considering the fact that a system prefers the state of lowest Gibbs free energy in the canonical ensemble, some locally thermodynamic stable black holes are globally unstable though they have positive heat capacity, such as states C and H. With the increase of the temperature, the black hole system evolves through the states A--B--E--E$'$--G--J and the horizon radius increases. At points E and E$'$, the system undergoes a small-large black hole phase transition, and a sudden change of black hole event horizon radius occurs.

Considering the off-shell Gibbs free energy $\GL$ as a function of the black hole event horizon, we investigate its behavior for five different temperatures $T_1-T_5$, respectively. For low temperature of the ensemble $T_1$, as shown in Fig.~\ref{Fig:off-shellb}, only one extremal point exists and it is located at the button of the well of the off shell Gibbs free energy. The extremal point corresponds to a stable black hole state A belonging to the small black hole branch. There are more black hole states with the increase of the temperature of the ensemble $T_{\rm E}$. When the temperature of the ensemble increases to $T_2$, different from the lower temperature $T_1$ case, three extremal points emerge as shown in Fig.~\ref{Fig:off-shellc} and they correspond to three black hole states. Two minimum points correspond to the locally stable small and large black hole states. The maximum point corresponds to the locally unstable intermediate black hole state. Increasing the temperature of the ensemble further, when it equals the small-large black hole phase transition temperature, i.e., $T_{\text{E}}=T_3$, the system undergoes a small-large black hole phase transition. Distinguished from other situations, the off shell Gibbs free energy exhibits a double-well behavior and the wells have the same depth as shown in Fig.~\ref{Fig:off-shelld}. The small and large black holes are the minimum points of the off shell Gibbs free energy and are located at the bottom of the wells, while the intermediate black hole state F is the maximum point and is located at the top of the barrier. In the viewpoint of the free energy landscape, the intermediate black hole state can be regarded as the potential barrier for the phase transition between the small black hole state E and the large black hole state E$'$~\cite{AnLe20}.  From the behavior of the off shell Gibbs free energy, it is obvious that the black hole horizon radius undergoes a sudden change for the small-large black hole phase transition. When the temperature of the ensemble $T_{\text{E}}=T_4$, there are three extremal points on the curve of the off shell Gibbs free energy and these points correspond to three black hole states. Contrary to the $T_{\text{E}}=T_2$ and $T_{\text{E}}=T_3$ cases, the Gibbs free energy of the large black hole is smaller, and the well is deeper on the off shell Gibbs free energy. Increasing the temperature of the ensemble to $T_{\text{E}}=T_5$, only one extremal point exists in the off shell Gibbs free energy and it corresponds to a stable large black hole state.

In summary, states A, B, E, E$'$, G and J are globally thermodynamic stable and correspond to the global minimum points of the off shell Gibbs free energy; states C and H are locally thermodynamic stable but globally unstable and are the local minimum points instead of global minimum points of the off shell Gibbs free energy. Intermediate black holes are local unstable and are located at the top of the barrier of the off shell Gibbs free energy. In particular, for the small-large black hole phase transition temperature, the off shell Gibbs free energy exhibits a double-well behavior and the wells have the same depth. While the intermediate black hole state acts as a barrier between the small and large black holes.

\section{Kinetic properties of phase transition}\label{Sec: Dynmic}

The free energy landscape shows the existence of states at some fixed temperatures and the possible transition between these states. In particular, on the small-large black hole phase transition temperature, the off shell Gibbs free energy exhibits the double well behavior and the wells have the same depth. In this section, we investigate the kinetic process of the stable small-large black hole phase transition using the Fokker-Planck equation.

\subsection{Fokker-Planck equation and probabilistic evolution}\label{Sec:FPEq}

Recently, Li and Wang initiated a proposal to study the kinetic process of black hole phase transition using the Fokker-Planck equation~\cite{LiWa20}. The Fokker-Planck equation is an equation of motion governing the distribution function of fluctuating macroscopic variables~\cite{Risk84}. For black hole phase transition, the event horizon radius $\rh$ is the order parameter, and it can be treated as a stochastic fluctuating variable during phase transition. Li and Wang proposed that it is possible to study the stochastic dynamics of black hole phase transition in terms of the associated probabilistic Fokker-Planck equation on the free energy landscape, and argued that the Fokker-Planck equation governs the probabilistic evolution of the order parameter for black hole phase transition~\cite{LiWa20}. We follow this proposal to investigate the kinetic process of the small-large black hole phase transition for the Kerr-AdS black hole.

We consider that there are a large number of black hole states in the thermodynamic ensemble with fixed temperature. The black hole states are described by the order parameter $\rh$. The probabilistic distribution of the black hole states evolving in time is a function of the order parameter. For the sake of simplicity, we denote the black hole event horizon $\rh$  by $r$ from now on. The probabilistic distribution function is denoted by $\rho(t,r)$.

On the free energy landscape, the off shell Gibbs free energy plays the role of an effective potential and acts as a driving force for black hole phase transition~\cite{LiZW21}, and it is a function of black hole horizon radius $r$, angular momentum $J$, thermodynamical pressure $P$ and temperature of the ensemble $T_{\rm E}$.  The evolution of the probabilistic distribution function $\rho(t,r)$ is governed by a special case of the Fokker-Planck equation known as the Smoluchowski equation~\cite{Zwan01},
\begin{equation}\label{SEequation}
  \frac{\partial \rho(t,r)}{\partial t}=D\frac{\partial}{\partial r}\left\{e^{-\beta\GL}\frac{\partial}{\partial r}\left[e^{\beta\GL}\rho(t,r)\right]\right\},
\end{equation}
where the diffusion coefficient $D=k_{\rm B}T_{\rm E}/\zeta $ and the parameter $\beta=1/(k_{\rm B}T_{\rm E}) $ with $k_{\rm B}$ and $\zeta $ being the Boltzmann constant and dissipation coefficient, respectively. The probabilistic distribution function $\rho (t,r)$ describes the probabilistic distribution of states after a thermodynamic fluctuation. Without loss of generality, we set $\zeta=1$ and use geometric units with $k_{\rm B}=1$.

To solve the Smoluchowski equation, we need to impose boundary conditions and the initial condition. There are two types of boundary conditions, one is the reflection boundary condition, which preserves the normalization of the probability distribution; the other is the absorption boundary condition.

Let us assume the boundary is located at $r_0$. {Expressed mathematically, the reflection boundary condition means that the probability current vanishes at the boundary~\cite{LiWa20},}
\begin{equation}\label{eq:reflection}
  j(t,r_0) = -De^{-\beta\GL}\frac{\partial}{\partial r}\left. \left[e^{\beta\GL}\rho(t,r)\right]\right| _{r=r_0}=0.
\end{equation}
While the absorption boundary condition means that the probability distribution function vanishes at the boundary,
\begin{equation}\label{eq:absorption}
  \rho(t,r_0)=0.
\end{equation}
Imposing which kind of boundary condition depends on the physical problem we consider.

In the present case of studying the time evolution of probability distribution of states in the canonical ensemble, the order parameter is the black hole event horizon, which ranges from the lower bound to infinity. The probability should be preserved throughout the evolution. We impose the reflection boundary condition at the lower bound $r=r_{\rm ex}$ and at infinity.

Since we are considering the kinetic process of the small-large black hole phase transition, the initial condition is chosen as the small black hole or large black hole state. We use Gaussian wave packet to describe the initial state with black hole event horizon $r=r_{\rm i}$,
\begin{equation}\label{eq:initial}
  \rho(0,r_{\rm i})=\frac{1}{\sqrt{\pi}\sigma}e^{-\frac{(r-r_{\rm i})^2}{\sigma^2}},
\end{equation}
where the initial state $r_{\rm i}$ is the small black hole state $r_{\rm s}$ or large black hole state $r_{\rm l}$ and the parameter $\sigma$ is set to $\sigma=0.1$.

\begin{figure*}[!htb]
  \begin{center}
\subfigure[]{\includegraphics[width=3.2in]{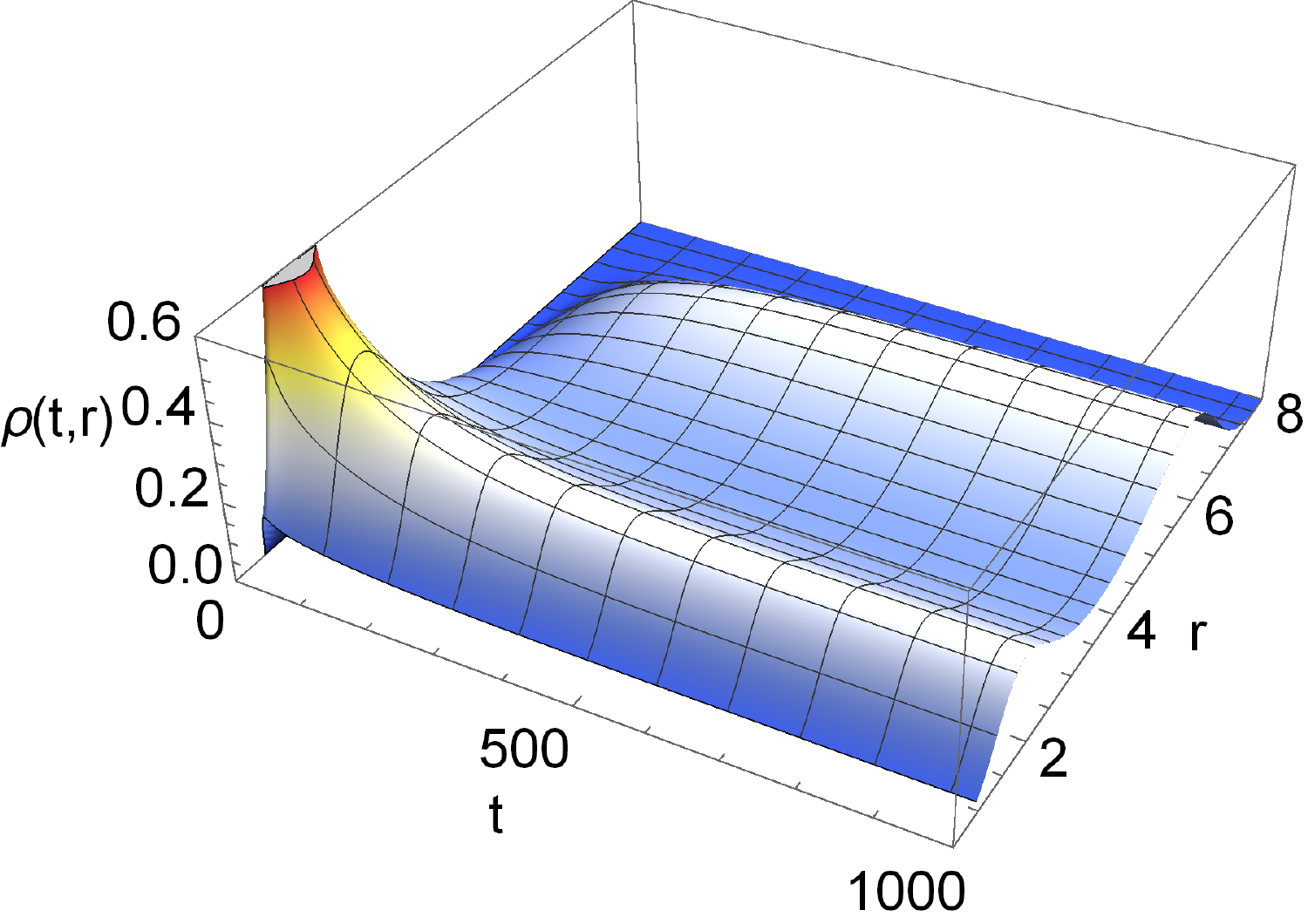}\label{Fig:distria}}
\subfigure[]{\includegraphics[width=3.2in]{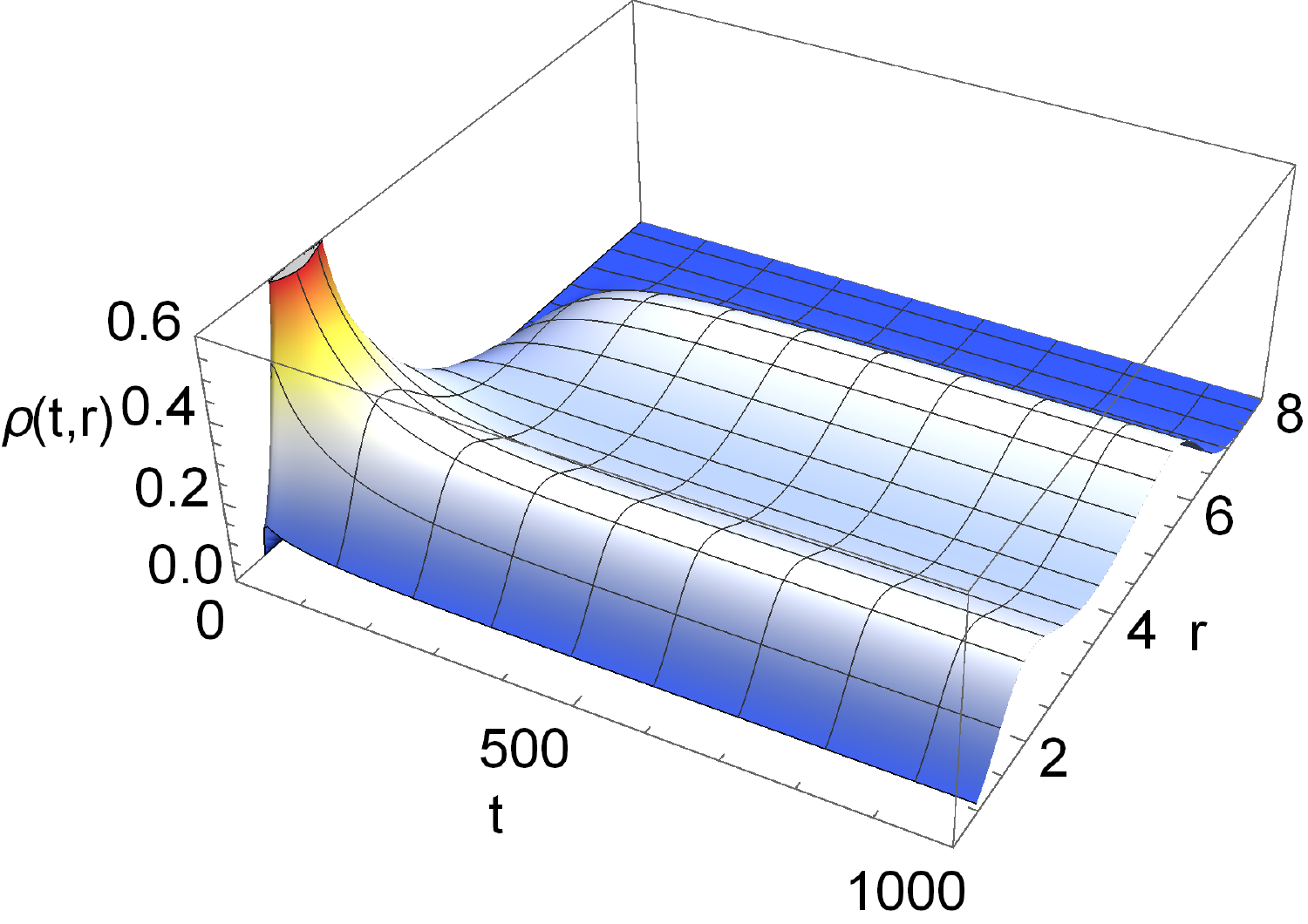}\label{Fig:distrib}}
\subfigure[]{\includegraphics[width=3.2in]{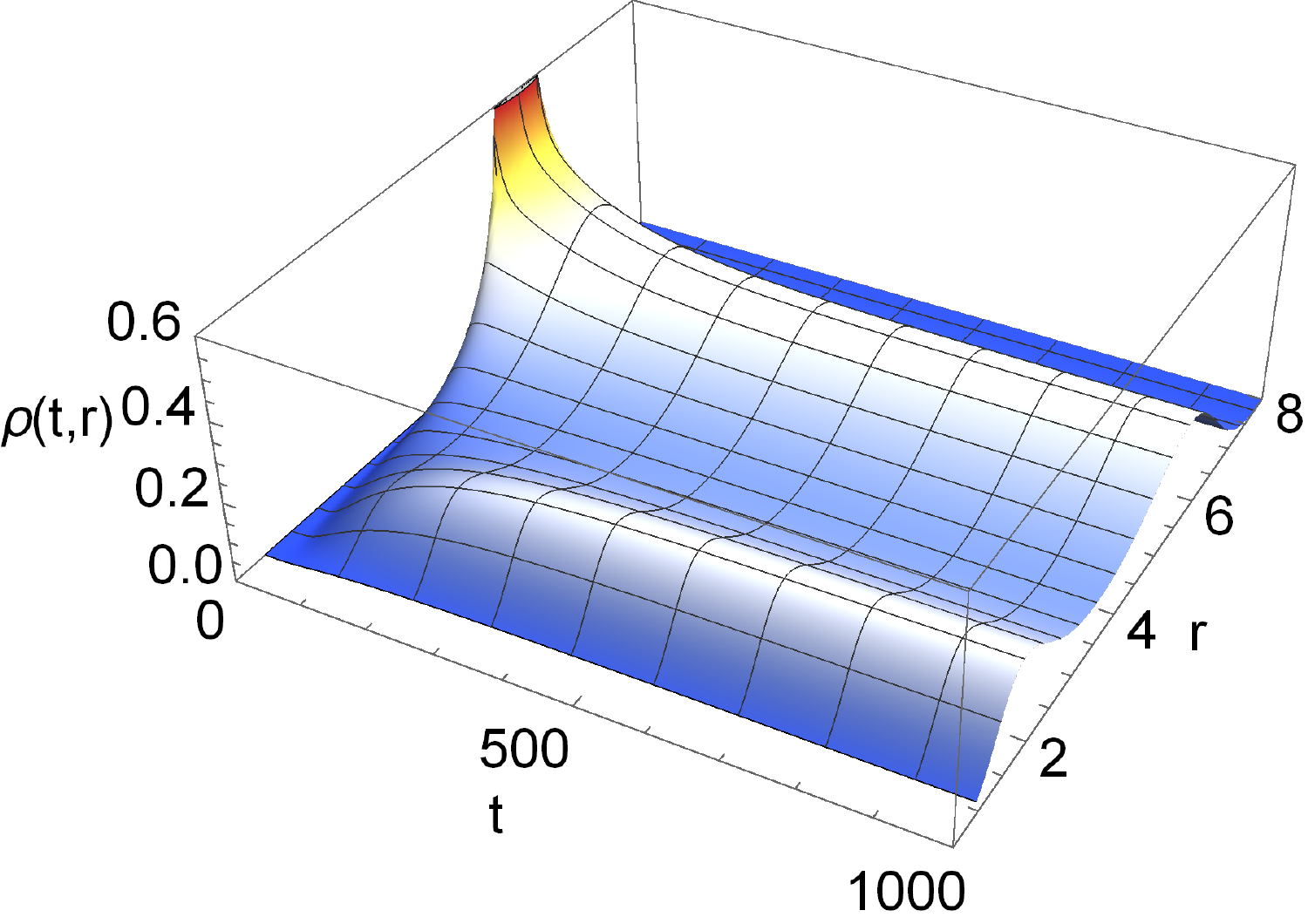}\label{Fig:distric}}
\subfigure[]{\includegraphics[width=3.2in]{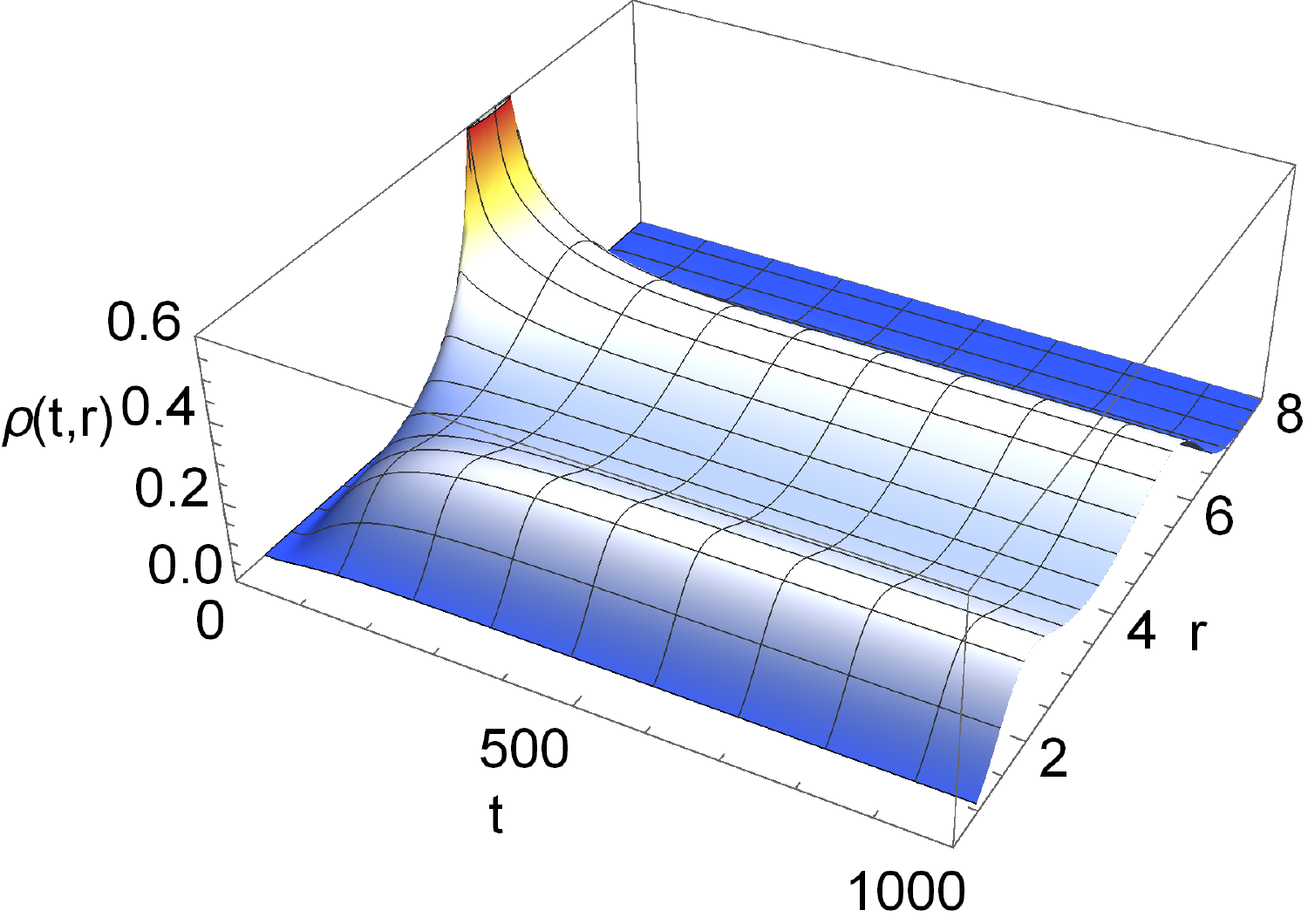}\label{Fig:distrid}}
  \caption{The evolution of the probability distribution function with $J=1$. In Fig.~\ref{Fig:distria} and Fig.~\ref{Fig:distric}, we set the temperature of the ensemble $T_{\rm E}=0.8306T_{\rm c}$ while the initial states are the small and large black hole, respectively. Then, we increase the temperature along the small-large black hole coexistence curve to $T_{\rm E}=0.8692T_{\rm c}$ in Fig.~\ref{Fig:distrib} and Fig.~\ref{Fig:distrid}.}\label{Fig:Distribution}
  \end{center}
\end{figure*}

The time evolution of the probability distribution is plotted in Fig.~\ref{Fig:Distribution}.
In the four figures, the temperatures and pressures correspond to points in the coexistence curve of phase transition. Since thermodynamics along the coexistence curve is extremely important, we consider the effect of the temperature on the small-large black hole phase transition by increasing the temperature along the coexistence curve. In Figs.~\ref{Fig:distria} and~\ref{Fig:distrib}, the initial states are the small black hole states but with $T_{\rm E}=0.8306T_{\rm c}$ and $T_{\rm E}=0.8692T_{\rm c}$, respectively. At $t=0$, the probability distribution is near the small black hole. The peak of the probability distribution $\rho (t,r)$ decreases with time, and the distribution probability for large black hole increases. This shows that there is a tendency for the small black hole to transit to the large black hole. The system evolves to a small-large black hole coexistence stationary state at a short time. The final stationary distribution of states is the Boltzmann distribution $\rho\propto e^{-\beta\GL(r)}$. The peaks are the small and large black hole states and they have the same probability. Figures~\ref{Fig:distric} and~\ref{Fig:distrid} are similar to Figs.~\ref{Fig:distria} and~\ref{Fig:distrib} but the initial states are the large black hole states.

\begin{figure*}[!htb]
  \begin{center}
\subfigure[]{\includegraphics[width=3.2in]{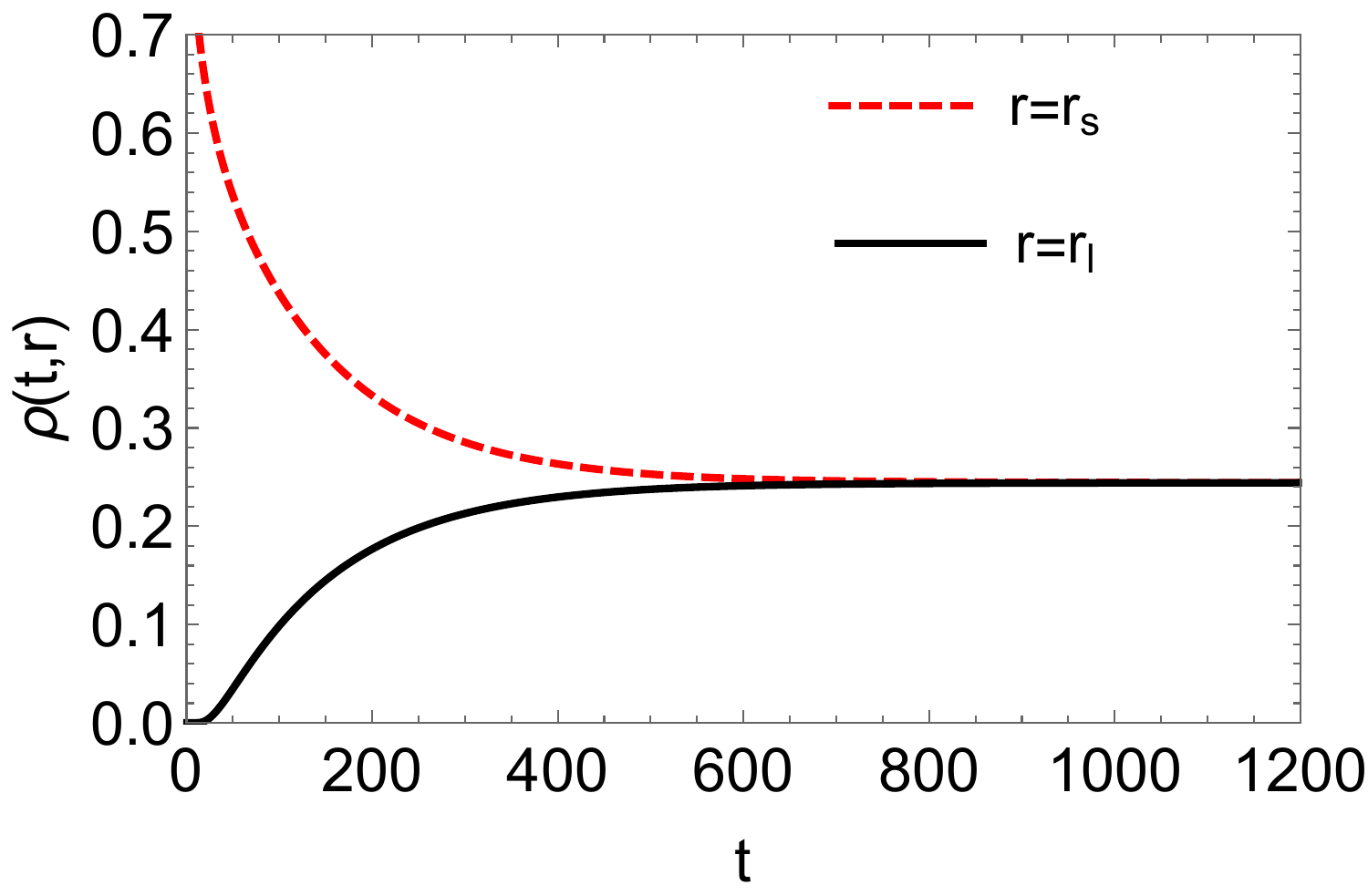}\label{Fig:Dista}}
\subfigure[]{\includegraphics[width=3.2in,height=2.05in]{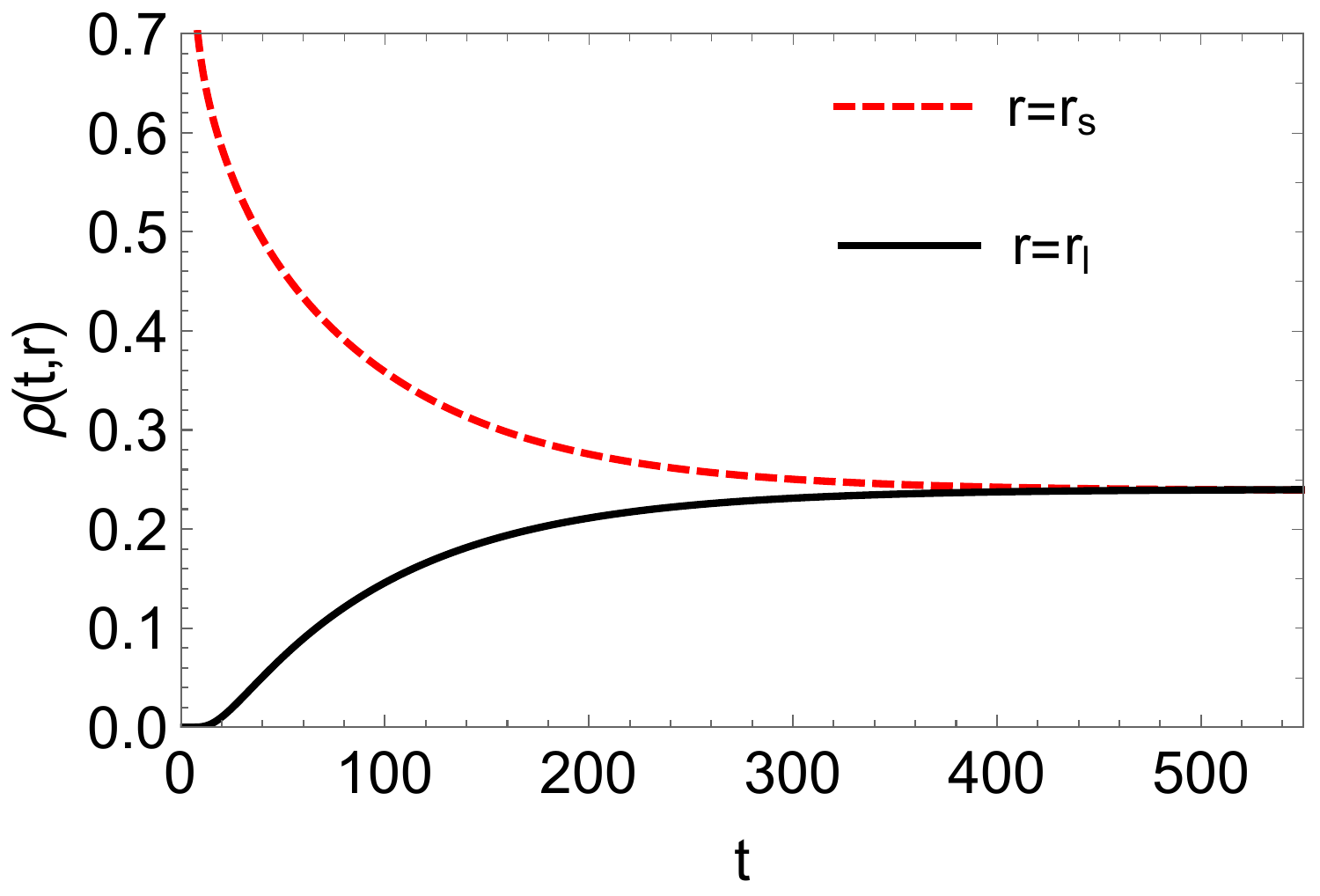}\label{Fig:Distb}}
\subfigure[]{\includegraphics[width=3.2in]{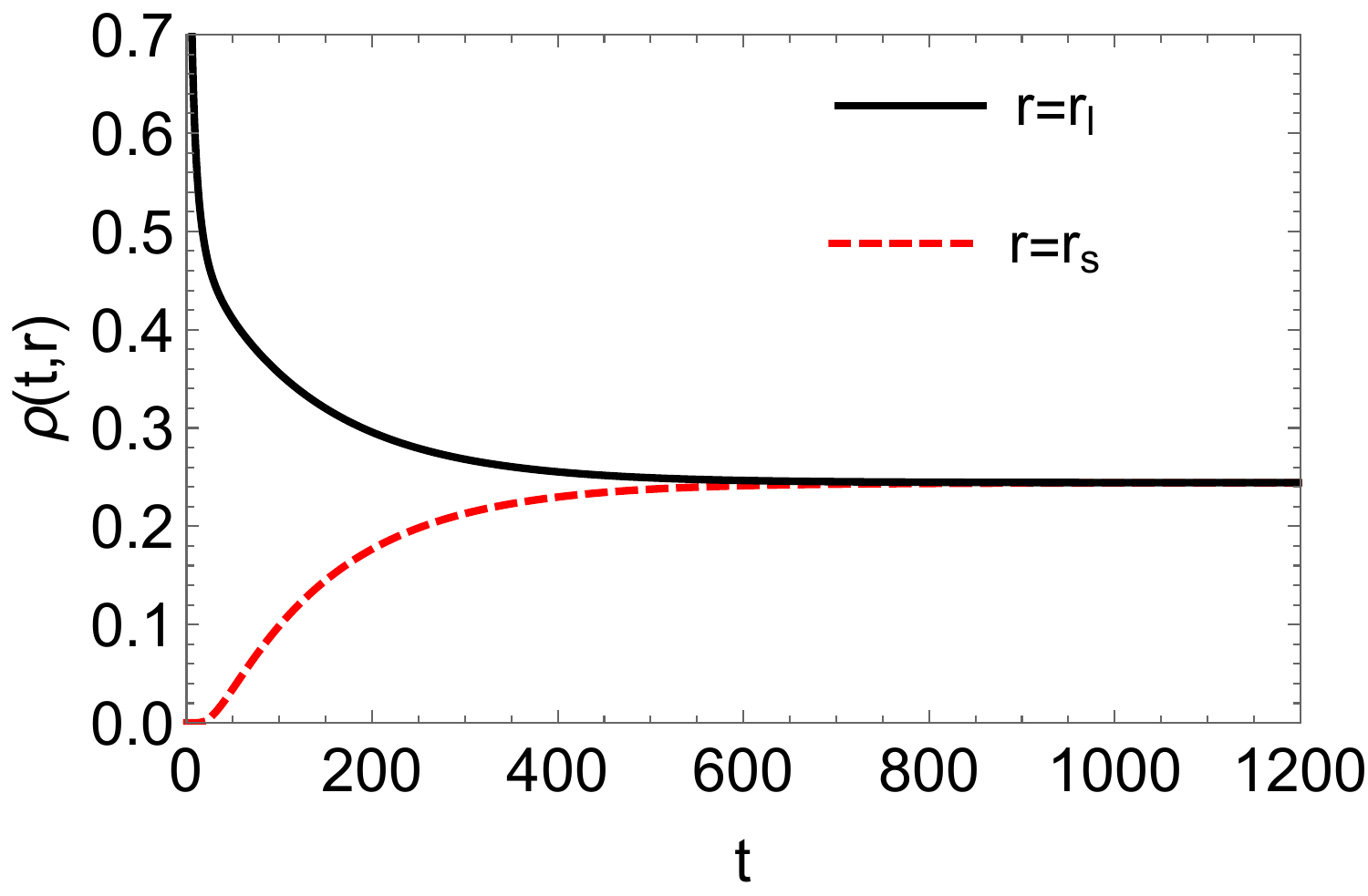}\label{Fig:Distc}}
\subfigure[]{\includegraphics[width=3.2in,height=2.05in]{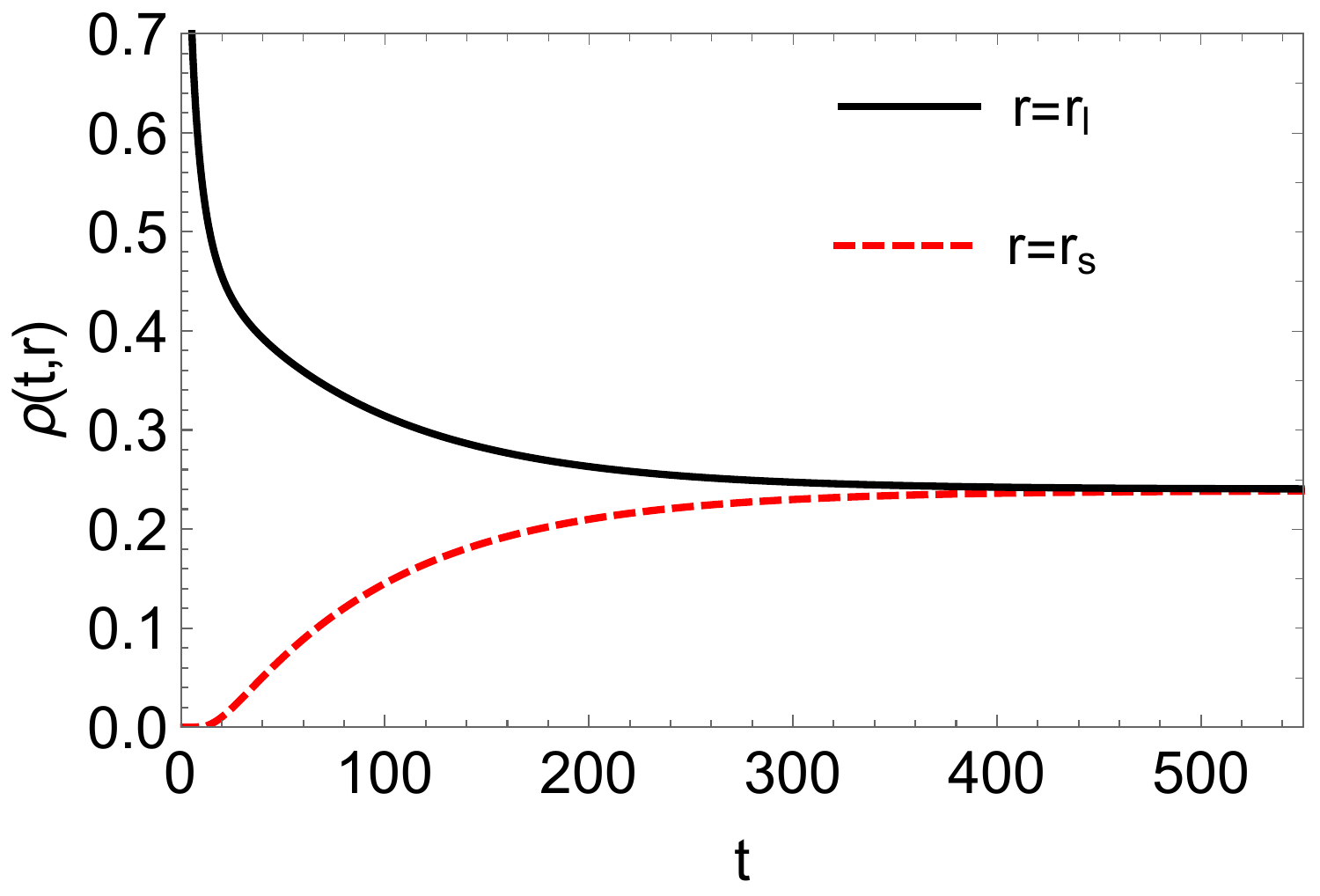}\label{Fig:Distd}}
  \caption{The evolution of the probability distribution function of the small and large black hole states with $J=1$. In Fig.~\ref{Fig:Dista} and Fig.~\ref{Fig:Distc}, the initial states are the small and large black holes, respectively. While the temperatures are $T_{\rm E}=0.8306T_{\rm c}$. Then, we increase the temperature of the ensemble along the small-large black hole coexistence curve to $T_{\rm E}=0.8692T_{\rm c}$. In Fig.~\ref{Fig:Distb} and Fig.~\ref{Fig:Distd}, the initial states are the small and large black holes, respectively, but the temperature are $T_{\rm E}=0.8692T_{\rm c}$.  }\label{Fig:Dist}
  \end{center}
\end{figure*}

To make the kinetic process for the small-large black hole phase transition clearer, we plot the probability distributions for the small black hole state $\rho(t,r_{\rm s})$ and large black hole state $\rho(t,r_{\rm l})$ in Fig.~\ref{Fig:Dist} 
for $T_{\rm E}=0.8306T_{\rm c}$ and $T_{\rm E}=0.8692T_{\rm c}$. In Fig.~\ref{Fig:Dista} and Fig.~\ref{Fig:Distb}, the initial states are the small black hole states $r_{\rm s}$; while in Fig.~\ref{Fig:Distc} and Fig.~\ref{Fig:Distd}, the initial states are the large black hole states $r_{\rm l}$. At $t=0$, the probability distribution for the small black hole state $\rho(t,r_{\rm s})$ takes finite value while the probability distribution for the large black hole state $\rho(t,r_{\rm l})$ vanishes in Fig.~\ref{Fig:Dista} and Fig.~\ref{Fig:Distb}. The probability distribution for the small black hole state $\rho(t,r_{\rm s})$ decreases while that of the large black hole state increases. This indicates that the small black hole states tend to transit to the large black hole states. The final probability distribution for the small black hole and large black hole states are the same since the wells of the off shell Gibbs free energy have the same depth for the small and large black hole states. Comparing Fig.~\ref{Fig:Dista} with Fig.~\ref{Fig:Distb}, it is clear that increasing the temperature of the ensemble reduces the time needed by the system to reach the final stationary distribution. This indicates that increasing the temperature makes the leaking for the small black hole state to the large one easier, since the barrier height for the small-large black hole phase transition is a monotonic decreasing function of the temperature. In Fig.~\ref{Fig:Distc} and Fig.~\ref{Fig:Distd} for the initial large black hole state, the results are similar.

As discussed in Sec.~\ref{Sec:thelandscape}, the intermediate black hole is located at the maximum point on the off shell Gibbs free energy, and it acts as the potential barrier for the small-large black hole phase transition. The barrier height will decrease with the increase of the temperature, and vanishes at the critical point $(P_{\rm c}, T_{\rm c})$. Hence, increasing the temperature of the ensemble along the coexistence curve makes that the initial stable small (large) black hole leaks to the stable large (small) black hole state more easily.

\subsection{The first passage time}\label{Sec:FPT}

In this subsection, we investigate the first passage time for the switching of the small and large black hole states. The first passage time is defined as the time that a stable small or large black hole state reaches the intermediate transition state located at the barrier of the off shell Gibbs free energy $\GL$ for the first time.

As discussed previously, the kinetic process of the small-large black hole phase transition is a stochastic process caused by thermal fluctuation, hence the first passage time is a random variable.  The distribution of the first passage time shows the timescale of the small-large black hole phase transition. We use $F_{\rm p}(t)$ to denote the distribution of the first passage time, and define $\Sigma(t)$ to be the probability that the black hole has not made a first passage by time $t$. The distribution of the first passage time $F_{\rm p}(t)$ and $\Sigma(t)$ are related by
\begin{equation}\label{Eq:FPT}
  F_{\rm p}(t)=-\frac{d\Sigma(t)}{dt}.
\end{equation}
It is clear that $F_{\rm p}(t)dt$ is the probability that the black hole has made its first passage in the time interval $(t, t+dt)$.

For the initial small black hole state, the probability that the black hole has not made the first passage by time $t$ is
\begin{equation}\label{Eq:SigmaS}
  \Sigma(t)=\int_{r_{\rm lb}}^{r_{\rm m}}\rho(t,r)dr,
\end{equation}
where $r_{\rm lb}$ and $r_{\rm m}$ are the lower bound of the order parameter and the horizon radius for the intermediate black hole, respectively.

Substituting Eq.~\eqref{Eq:SigmaS} into Eq.~\eqref{Eq:FPT} and imposing the reflection boundary condition at the lower bound $r_{\rm lb}$ and the absorption boundary condition at the intermediate black hole state $r_{\rm m}$ for the Fokker-Planck equation, we get
\begin{equation}
\begin{split}
    F_{\rm p}(t) &= -\frac{d}{dt}\int_{r_{\rm lb}}^{r_{\rm m}}\rho(t,r)dr \\
   &= -\int_{r_{\rm lb}}^{r_{\rm m}}\frac{\partial \rho(t,r)}{\partial t}dr \\
   &= -\int_{r_{\rm lb}}^{r_{\rm m}}D \frac{\partial}{\partial r}\left\{e^{-\beta \GL(r)}\frac{\partial}{\partial r}\left[e^{\beta \GL(r)}\rho(t,r)\right]\right\}dr  \\
     &=-D\left. e^{-\beta \GL(r)}\frac{\partial}{\partial r}\left[e^{\beta \GL(r)}\rho(t,r)\right]\right|_{r_{\rm lb}}^{r_{\rm m}} \\
    & = -D\left. \frac{\partial \rho(t,r)}{\partial r} \right|_{r=r_{\rm m}}.
\end{split}
\end{equation}

With similar procedure for the initial large black hole state but imposing the absorption boundary condition at the intermediate black hole state $r_{\rm m}$ and the reflection boundary condition at infinity for the Fokker-Planck equation, the probability $\Sigma(t)$ that the initial large black hole has not made a first passage by time $t$ and the distribution of the first passage time $F_{\rm p}(t)$ are
\begin{gather}
  \Sigma(t) = \int^{+\infty}_{r_{\rm m}}\rho(t,r)dr, \\
  F_{\rm p}(t)= D\left.\frac{\partial \rho(t,r)}{\partial r}\right|_{r=r_{\rm m}}.
\end{gather}

\begin{figure*}[!htb]
  \begin{center}
\subfigure[]{\includegraphics[width=3.2in]{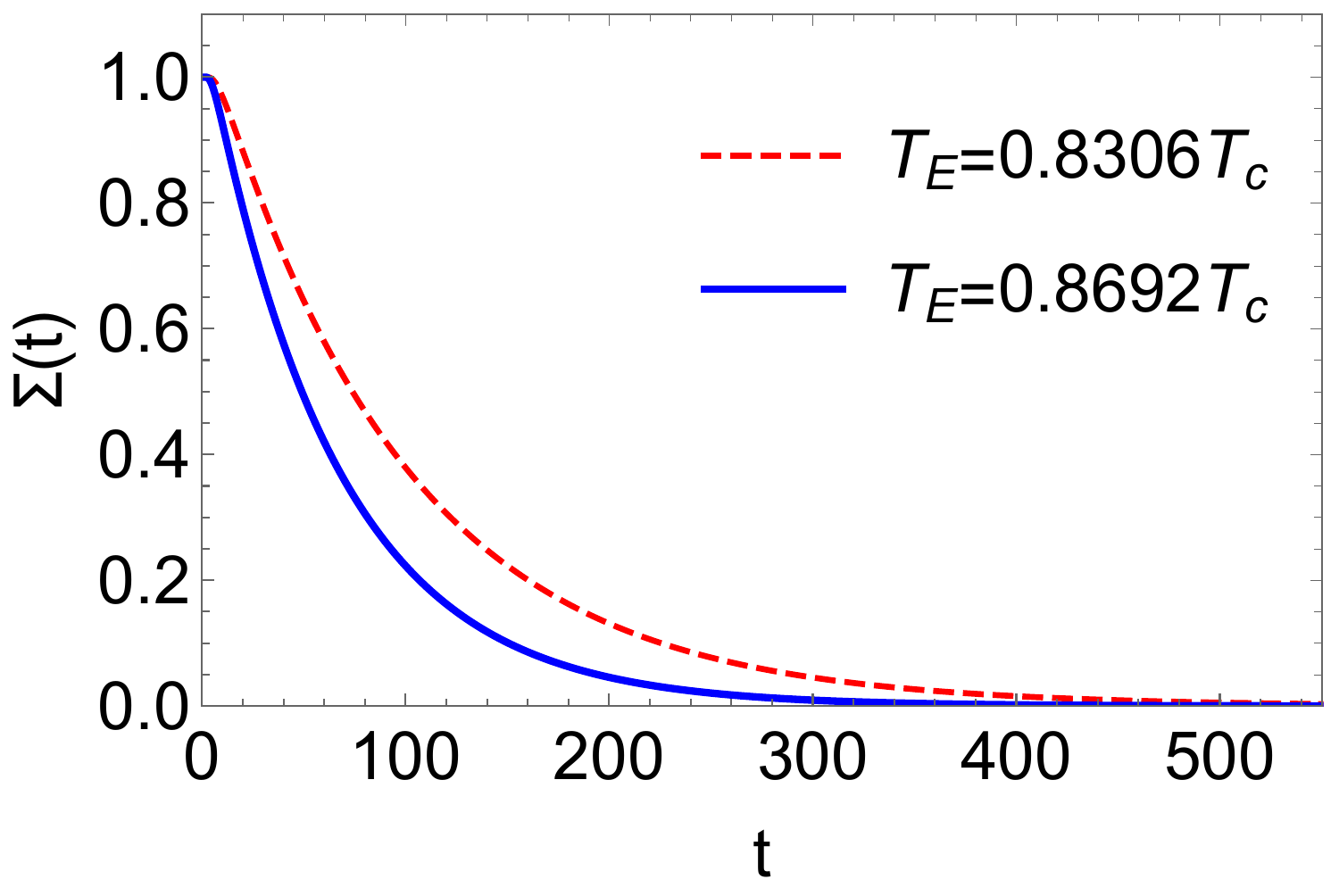}\label{Fig:SigmaS}}
\subfigure[]{\includegraphics[width=3.2in]{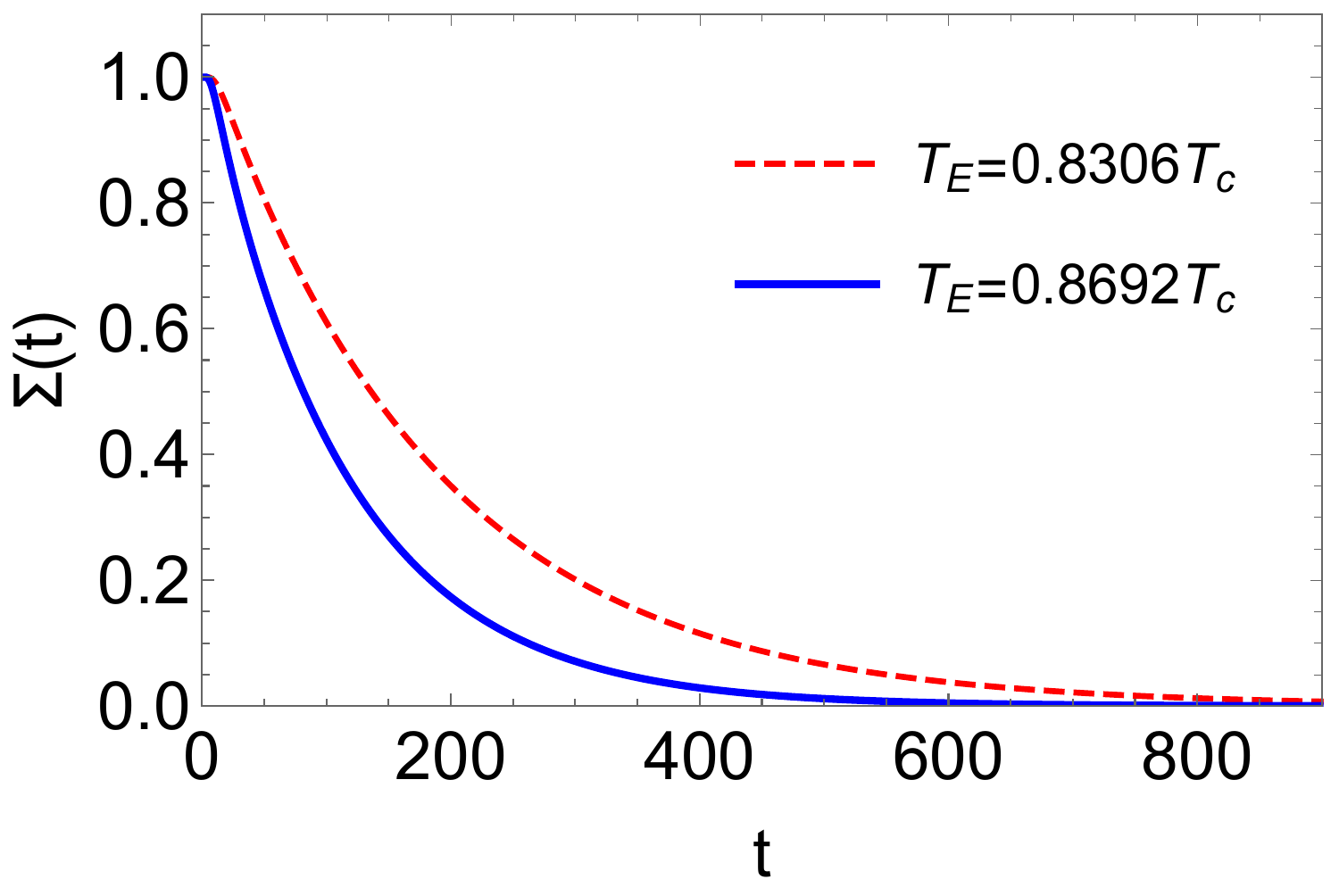}\label{Fig:SigmaL}}
  \caption{The evolution of the probability that the initial black hole has not made a first passage process with $J=1$. (a).~The initial state is a small black hole. (b).~The initial state is a large black hole.  }\label{Fig:SigmaDist}
  \end{center}
\end{figure*}

To show the first passage process precisely, we depict the probability $\Sigma(t)$ that the initial black hole state has not made its first passage by time $t$ for initial small and large black hole states with temperatures $T_{\rm E}=0.8306T_{\rm c}$ and $T_{\rm E}=0.8692T_{\rm c}$ in Fig.~\ref{Fig:SigmaDist}.
From the figures, it is obvious that the initial black hole state decays quickly and the probability decreases more quickly with the increase of the temperature.

\begin{figure*}[!htb]
  \begin{center}
\subfigure[]{\includegraphics[width=3.2in]{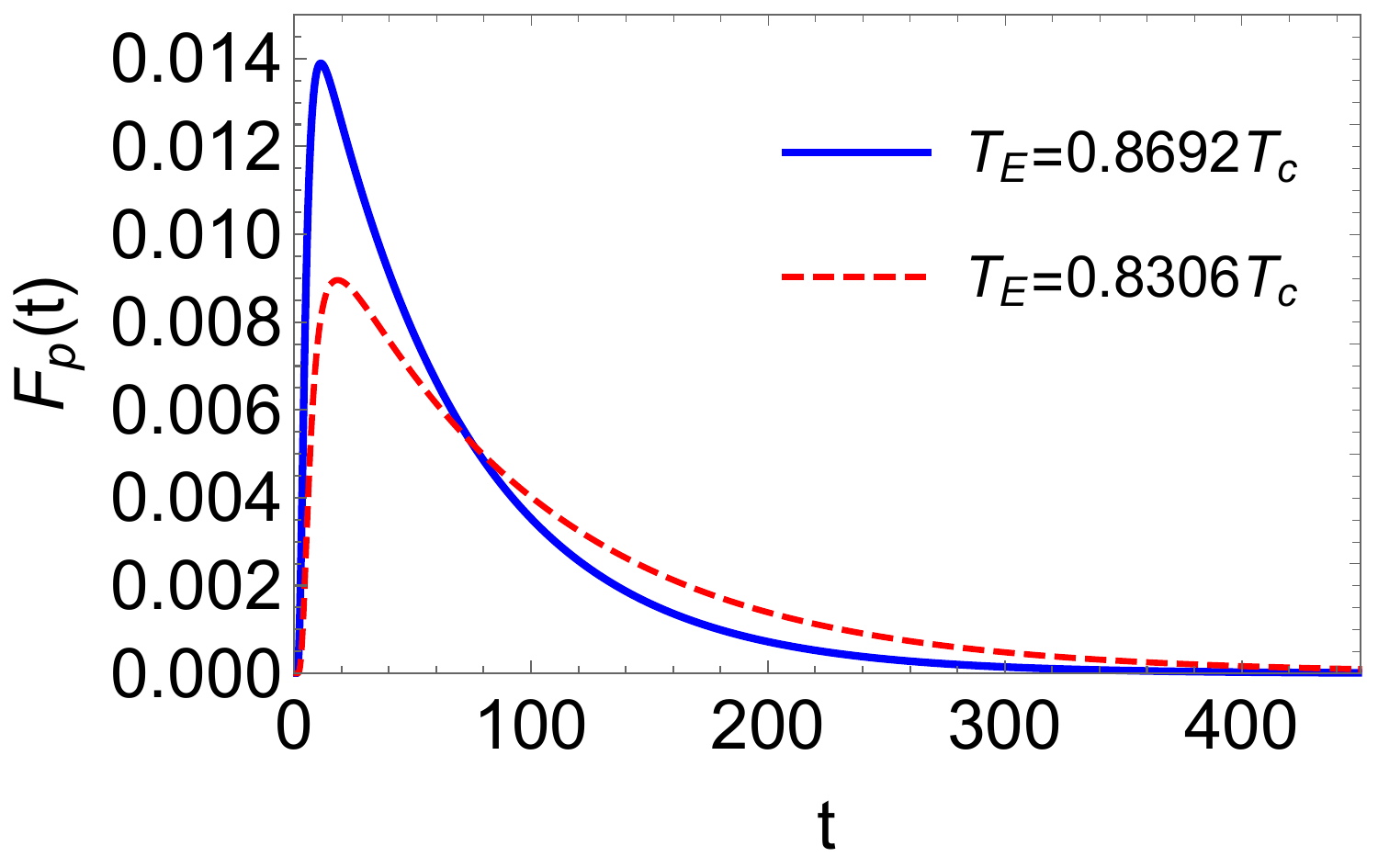}\label{Fig:FPTSBH}}
\subfigure[]{\includegraphics[width=3.2in]{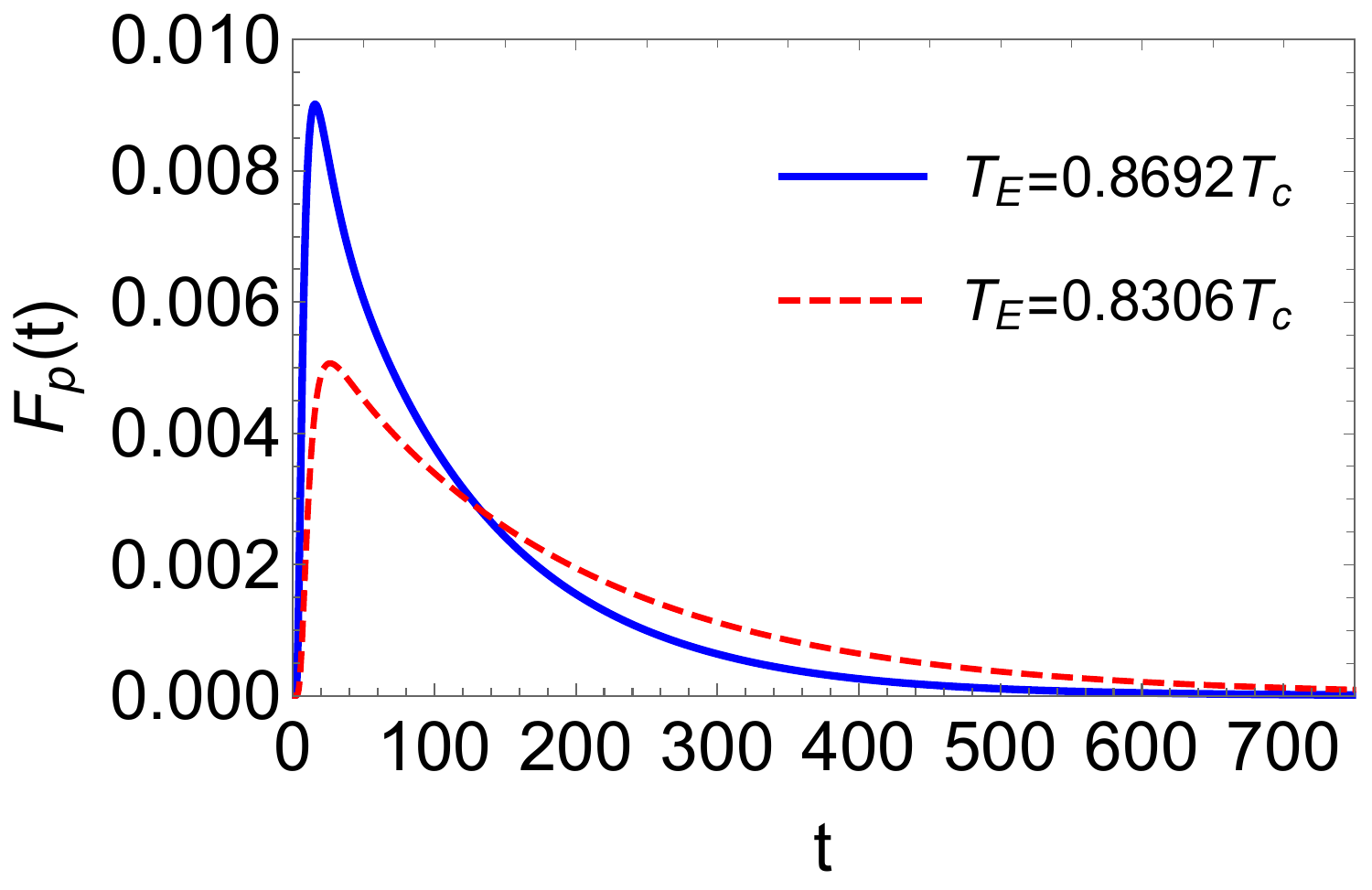}\label{Fig:FPTLBH}}
  \caption{The figure for the distribution of the first passage time for initial small and large black hole with different temperatures. Where we set the angular momentum of the black hole $J=1$. (a).~The distribution of the first passage time for initial small black hole state. (b).~The distribution of the first passage time for initial large black hole state. }\label{Fig:FPTDist}
  \end{center}
\end{figure*}

The distribution for the first passage time for initial small and large black hole states is displayed in Fig.~\ref{Fig:FPTDist}.
The behaviors of the initial small black hole state in Fig.~\ref{Fig:FPTSBH} and the initial large black hole state in Fig.~\ref{Fig:FPTLBH} are similar. There is a peak for each curve. The distribution for the first passage time increases quickly and reaches the peak at a short time. Then the distribution for the first passage time decreases quickly. This means that a considerable first passage process occurs at a short time. The increase of the temperature makes the peak sharper and the peak shifts to the left. This suggests that increasing the temperature of the ensemble makes the first passage process easier to occur due to the decrease of barrier height as discussed in Sec.~\ref{Sec:FPEq}.

\section{Discussion and conclusion}\label{Sec:Conclu}

Thermodynamics and phase transition for black holes provide us a perfect test bed for quantum gravity. Furthermore, the study of black hole thermodynamics and phase transition might shed light on the underlying microscopic structure of black hole. The exploration of the kinetics of black hole phase transition might provide insight into the underlying microscopic structure of black hole. In this paper, we investigated the dynamics and kinetics of the small-large black hole phase transition for the Kerr-AdS black hole on free energy landscape.

First, we reviewed the thermodynamics and coexistence curve for the Kerr-AdS black hole. Due to a sudden change of the black hole event horizon for small-large black hole phase transition below the critical temperature and the change becomes zero at the critical point, we regarded the black hole event horizon as the order parameter. Then we analyzed the stability and the small-large black hole phase transition via the heat capacity and the Gibbs free energy. To investigate the kinetics of the small-large black hole phase transition, we generalized the Gibbs free energy to off shell Gibbs free energy. The off shell Gibbs free energy can be regarded as a function of the order parameter $\rh$ and we proved that the extremal points of the off shell Gibbs free energy $\GL(r_{\rm h})$ for the Kerr-AdS black hole correspond to physical black holes. In particular, we found that there is a lower bound for the order parameter for the Kerr-Newman-AdS family black holes. The lower bound corresponds to extremal black holes. We analytically solved the lower bound of the order parameter for the Kerr-AdS black hole. Moreover, we found that the off shell Gibbs free energy approaches zero instead of divergent as previous work suggested when the black hole event horizon approaches zero. Our work is consistent with the physical intuition that vanishing black hole event horizon corresponds to a thermal AdS space, and the off shell Gibbs free energy for the thermal AdS space is zero. Having the off shell Gibbs free energy, we analyzed the behavior of the off shell Gibbs free energy for various temperatures. We found that there is a double well behavior for stable small-large black hole phase transition and the wells have the same depth.

Since the stable small-large black hole phase transition is a stochastic thermal fluctuation of the order parameter, and the Fokker-Planck equation governs the distribution of fluctuating macroscopic variables, we use the recently initiated proposal that the kinetics of black hole phase transition can be described by the Fokker-Planck equation to explore the dynamical behavior of the small-large black hole phase transition. We solved the Fokker-Planck equation numerically for initially small and large black hole states, respectively. The results suggest that the initial stable small (large) black hole states tend to fluctuate to stable large (small) black hole states. Increasing the temperature along the coexistence curve, the switching process becomes quicker and reaches the final stationary Boltzmann distribution at a shorter time. We then analyzed the first passage process for the small-large black hole phase transition. The distribution of the first passage time shows the timescale of the small-large black hole phase transition. We found that there is a peak for the distribution of the first passage time. This indicates that a considerable first passage process occurs at a short time. Moreover, the peak in the distribution of the first passage time becomes sharper and shifts to the left with the increase of the temperature along the coexistence curve. This demonstrates that a considerable first passage process occurs at a shorter time for a higher temperature.

Thermodynamics of black holes along the coexistence curve is an intriguing subject. The investigation of the kinetic process of black hole phase transition along the coexistence curve might lead to the progress and understanding of black hole microscopic structures.

\acknowledgments

We thank Wen-Di Guo, Jun-Jie Wan, Zi-Chao Lin, Jing Chen, Zheng-Quan Cui, and Tao-Tao Sui for many useful discussions. This work was supported in part by the National Natural Science Foundation of China (Grants No. 11875151, No. 12075103, and No. 12047501), the Fundamental Research Funds for the Central Universities (Grants No.~lzujbky-2021-it34, No.~lzujbky-2019-it21, and No.~lzujbky-2019-ct06), and the 111 Project (Grant No. B20063).


\end{document}